\documentclass[11pt]{article}
\usepackage{amssymb}
\usepackage[dvips]{graphicx}
\usepackage[usenames]{color}
\usepackage{ulem}
\usepackage{amsmath}
\begin{document}
\begin{center}
{\large\bf Dynamical System of Scalar Field from 2-Dimension to 3-D
and its Cosmological Implications}

 \vskip 0.15 in
 $^\dag$Wei Fang$^{1, 2, 3}$, Hong Tu$^{1, 2}$, Jiasheng Huang$^3$, Chenggang Shu$^2$
 \\ \

\small {$^1$\textit{Department~of~Physics,~Shanghai~Normal~University,~100~Guilin~Rd.,~Shanghai,~200234,~P.R.China}\\
        $^2$\textit{The~Shanghai~Key~Lab~for~Astrophysics,~100~Guilin~Rd.,~Shanghai,~200234,~P.R.China}\\
        $^3$\textit{Harvard-Smithsonian~Center~for~Astrophysics,~60~Garden~St.,~Cambridge,~MA~02138,~USA}}

\footnotetext{$\dag$ \ \ wfang@shnu.edu.cn, wfang@cfa.harvard.edu}

 \vskip 0.5
in  \centerline{\bf Abstract} \vskip 0.2 in
\begin{minipage} {5.8in} {\hspace*{10pt}\small
\\We give the three-dimensional dynamical autonomous systems for most of
the popular scalar field dark energy models including (phantom)
quintessence, (phantom) tachyon, k-essence and general non-canonical
scalar field models, change the dynamical variables from variables
$(x, y, \lambda)$ to observable related variables $(w_{\phi},
\Omega_{\phi}, \lambda)$, and show the intimate relationships
between those scalar fields that the three-dimensional system of
k-essence can reduce to (phantom) tachyon, general non-canonical
scalar field can reduce to (phantom) quintessence and k-essence can
also reduce to (phantom) quintessence for some special cases. For
the applications of the three-dimensional dynamical systems, we
investigate several special cases and give the exactly dynamical
solutions in detail. In the end of this paper, we argue that, it is
more convenient and also has more physical meaning to express the
differential equations of dynamical systems in  $(w_{\phi},
\Omega_{\phi}, \lambda)$ instead of variables $(x, y, \lambda)$ and
to investigate the dynamical system in 3-Dimension instead of
2-Dimension. We also raise a question about the possibility of the
chaotic behavior in the spatially flat single scalar field FRW
cosmological models in the presence of ordinary matter.
\\

 }
\end{minipage}
\end{center}

\newpage

\section{Introduction}
Scalar field models have played a vital role in cosmological
theoretical studies in nearly half a century. Those assumed scalar
fields appeared in different cosmological research aspects  to
settle different cosmological problems \cite{1}, such as to drive
inflation, to explain a time variable cosmological $'$constant$'$
and so on. After the discovery of the accelerating expansion of
universe, scalar fields have played another important essential role
as a candidate of dark energy . There are so many phenomenological
dark energy models of scalar fields, such as quintessence, phantom,
quintom and the scalar fields with non-canonical kinetic energy term
(for a review, see\cite{2,3}).

\par Phase-plane analysis is a very useful and common method(see Ref\cite{add2} and recent papers,
e.g. \cite{Saridakis1, Saridakis2, Saridakis3, Saridakis4}) to study
the dynamical evolution of those scalar fields models and their
cosmological implications. However, most of those works only focus
on the quintessence models(including phantom quintessence and
quintom) with unique exponential potential and tachyon
models(including phantom tachyon) with inverse square potential, and
correspondingly, the dynamical systems are two dimensional
autonomous system(see the references cited in \cite{6, 12}). Using a
method which considers the potential related variable $\Gamma$ as a
function of another potential related variable $\lambda$ (see
Eq.(\ref{eqsadd4}) for the definition of $\Gamma$ and
$\lambda$)\cite{26, 6, 12}, we are able to analyze the phase-plane
of the dynamical systems of the quintessence and tachyon models with
many different potentials. When the potentials are beyond the
special type such as exponential or inverse square potentials, the
dynamical systems consequently become a three-dimensional autonomous
systems. This method is quite effective and powerful, it therefore
has been generalized to several other cosmological
contexts\cite{cite61, cite62, cite63, cite65, cite66, cite67,
cite68, cite69, 25, cite611}. However, there is very few work
focusing on the dynamical behavior of the scalar field with a
general modified kinetic term, such as k-essence ($L=V(\phi)F(X)$)
and general non-canonical scalar field ($L=F(X)-V(\phi)$). Recently,
Josue De-Santiago et. al analyzed the dynamical system of general
non-canonical scalar field with the lagrangian $L=F(X)-V(\phi)$ and
studied the phase plane after a suitable choice of
variables\cite{20}. They obtained the three-dimensional autonomous
system of this non-canonical scalar field after specifying the
kinetic term as $F(X)=AX^{\eta}$ and choosing the potential as
$V(\phi)=V_0(\phi-\phi_0)^{1/(1-\Gamma)}$(i.e., a special case that
$\Gamma=const$) and studied the critical points as well as their
stability.
\par Motivated by the work\cite{20}, we try to extend our works\cite{6, 12} in this paper to
give the three-dimensional autonomous dynamical systems for most of
the popular scalar field dark energy models including (phantom)
quintessence, (phantom) tachyon, k-essence and general non-canonical
scalar field models in current work. We will show that the
three-dimensional autonomous systems of general non-canonical scalar
field and k-essence will reduce to the quintessence and tachyon
scalar field respectively. Not like most of the previous works, here
we express the three dimensional autonomous systems from variables
$(x, y, \lambda)$ to the observable related variables $(w_{\phi},
\Omega_{\phi}, \lambda)$. It will be very convenient to investigate
the dynamical properties of the autonomous system based on the
observable related variables $w_{\phi}$ and $\Omega_{\phi}$(see
\cite{add2, add3, 9, 7, 8, 10, 12, 13} and a recent paper about the
general property of dynamical quintessence field\cite{add1}). Since
the definition of the variables $x$ and $y$ could vary with
different scalar field models, while the meaning of $w_{\phi}$ and
$\Omega_{\phi}$ are the same for different dark energy models and
therefore the differential equations of $w_{\phi}$ and
$\Omega_{\phi}$ are model independent. The paper is organized as
follows. We firstly present the basic theoretical framework for
(phantom) quintessence, (phantom) tachyon, k-essence and general
non-canonical scalar field models in section 2, and try to give the
relationships between those different scalar fields in this section.
We then give the three dimensional autonomous dynamical systems for
those scalar fields and switch the dynamical variables from $(x, y)$
to $(w_{\phi}, \Omega_{\phi})$ in each subsection of section 3.
Additionally, using the dynamical systems, we give the exact
solution of $w_{\phi}$ and $\Omega_{\phi}$ for a special case of
tachyon model when the potential is chosen to be a constant in
subsection 3.2. We show that the dynamical autonomous system of
k-essence can reduce to tachyon model, and investigate another
special case called kinetically driven quintessence with the
lagrangian $p(X, \phi)=f(\phi)(-X+X^2)$ detailedly in subsection
3.3. In subsection 3.4, we show that the dynamical autonomous system
of general non-canonical scalar field can reduce to quintessence and
tachyon model respectively for some special cases. We also studied
two special cases of purely kinetic united model $L = F(X)$ in
detailed in this subsection. We try to give the cosmological
implications of the three-dimensional dynamical autonomous system
and present the conclusion in section 4. We also raise a question
about the possibility of the chaotic behavior in the spatially flat
single scalar field FRW cosmological models in the presence of the
ordinary matter.

\section{Basic Framework for various Scalar Fields}

Let us restrict ourselves to a flat universe described by the FRW
metric, and consider a spatially homogeneous real scalar field
$\phi$ with non-canonical kinetic energy term. The lagrangian
density is given as

\begin{equation}\label{eqs1}L=L(X,V)\end{equation}

where $L$ is a function of $X$ and potential $V(\phi)$,
$X=\frac{1}{2}\nabla_{\mu}\phi\nabla^{\mu}\phi=\frac{1}{2}{\dot{\phi}}^2
$ for a spatially homogeneous scalar field. The pressure, energy
density and the Friedmann equations of the scalar field could be
easily obtained as following:

\begin{equation}\label{eqs2}p_{\phi}=L(X, V)\end{equation}
\begin{equation}\label{eqs3}\rho_{\phi}=2X L_X -L\end{equation}
\begin{equation}\label{eqs4}H^2=(\frac{\dot{a}}{a})^2=\frac{1}{3M^2_{pl}}[\rho_{\phi}+\rho_b]\end{equation}
\begin{equation}\label{eqsadd1}\dot H=-\frac{1}{2M^2_{pl}}[2X L_X+\gamma_b\rho_b]\end{equation}

where $8\pi G=\kappa^2=1/M^2_{pl}$, $\rho_b$ is the density of a
barotropic fluid component with the equation of state $p_b=w_b
\rho_b=(\gamma_b-1) \rho_b$. $\gamma_b=1$ for matter and
$\gamma_b=4/3$ for radiation. $L_X$ is the derivative of $L(X,\phi)$
with respect to $X$.
\par For quintessence, general non-canonical scalar field, tachyon and K-essence model, the pressure $p$ and energy density $\rho$ are respectively:

\begin{equation}\label{eqs5} p_q=\varsigma \frac{1}{2}\dot{\phi}^2-V(\phi), \rho_q=\varsigma \frac{1}{2}\dot{\phi}^2+V(\phi)\end{equation}
\begin{equation}\label{eqs8} p_g=F(X)-V(\phi),\rho_g=2XF_X-F(X)+V(\phi) \end{equation}
\begin{equation}\label{eqs6} p_t=-V(\phi)\sqrt{1-\varsigma \dot \phi^2}, \rho_t=\frac{V(\phi)}{\sqrt{1-\varsigma \dot \phi^2}}\end{equation}
\begin{equation}\label{eqs7} p_K=-V(\phi)F(X), \rho_K=2L_XX-L(X, \phi)=V(\phi)(F-2XF_X)\end{equation}

\par If $\varsigma=1$, Eq.(\ref{eqs5}) and Eq.(\ref{eqs6}) correspond to the quintessence and tachyon scalar field. If $\varsigma=-1$, Eq.(\ref{eqs5}) and
 Eq.(\ref{eqs6}) correspond to the phantom quintessence and phantom tachyon scalar field. General non-canonical scalar field Eq.(\ref{eqs8})
 can recover to (phantom) quintessence Eq.(\ref{eqs5}) if $F(X)=\varsigma {\dot\phi}^2=2\varsigma X$. K-essence model Eq.(\ref{eqs7}) can recover to (phantom) tachyon
 Eq.(\ref{eqs6}) if $F(X)=\sqrt{1-\varsigma \dot\phi^2}=\sqrt{1-2\varsigma X}$. Moreover, if the scalar field is redefined, it is demonstrated that K-essence model described by Eq.(\ref{eqs7}) with a linear kinetic function $F(X)=X+1$
 can reduce to any quintessence model described by Eq.(\ref{eqs5}). It means that any quintessence can be contained into K-essence frame, so each quintessence model
 is kinematically equivalent to a k-essence model.  The authors also give the relationship between the potentials of the two models\cite{4}. For example,
 the exponential potential $V(\phi)=V_0 e^{-\lambda\phi}$
 in quintessence model plays the similar role as the inverse square potential $V(\phi)=(\frac{1}{2}\kappa \lambda \phi-c_1)^{-2}$. We will also show
 that the role of inverse square potential in k-essence model is very similar with exponential potential in quintessence in the next section.

\section{Dynamical System of various Scalar Fields}
\par In this section, we will give the dynamical system for all the quintessence, tachyon, K-essence and general non-canonical scalar field model. We will
 summarize the dynamical system analysis and give our comments.
\subsection{Dynamical System for Quintessence and Phantom Quintessence Scalar Field}
\par For the (phantom) quintessence scalar field with lagrangian $L=\varsigma \frac{1}{2}\dot{\phi}^2-V(\phi)$, we can define the
following dimensionless variables:

\begin{equation}\label{eqsadd4} x=\frac{\kappa\dot\phi}{\sqrt6 H}, y=\frac{\kappa\sqrt{V}}{\sqrt3 H},~ \lambda(\phi)=-\frac{V'}{V}, \Gamma(\phi)=\frac{V V''}{V'^2}\end{equation}
 Where $V'=dV(\phi)/d\phi, V''=d^2V(\phi)/d\phi^2 $. The parameters $\Gamma(\phi)$ and $\lambda(\phi)$ of the potentials can be related with the famous
 slow roll parameters $\epsilon_V$ and $\eta_V$(e.g., see \cite{5}):

\begin{equation}\label{eqsadd2}\epsilon_V(\phi)=\frac{M^2_{pl}}{2}(\frac{V'}{V})^2=\frac{M^2_{pl}}{2}\lambda^2(\phi)\end{equation}
\begin{equation}\label{eqsadd3}\eta_V(\phi)=M^2_{pl}\frac{V''}{V}=M^2_{pl}\frac{V}{V'^2}\frac{V''}{V}\frac{V'^2}{V}=M^2_{pl}\Gamma(\phi)\lambda^2(\phi)\end{equation}

 Using Eq.(\ref{eqs4}), Eq.(\ref{eqsadd1}),Eq.(\ref{eqs5})and Eq.(\ref{eqs8}), We can write down the following equations for the evolution of the
 (phantom) quintessence:

\begin{equation}\label{eqs9}\frac{dx}{dN}=-3x+\frac{\sqrt{6}}{2}\varsigma \lambda y^2+\frac{3}{2} x[\varsigma (1-w_b)x^2+(1+w_b)(1-y^2)]\end{equation}
\begin{equation}\label{eqs10}\frac{dy}{dN}=-\frac{\sqrt{6}}{2}\lambda xy+\frac{3}{2}y[\varsigma (1-w_b)x^2+(1+w_b)(1-y^2)]\end{equation}
\begin{equation}\label{eqs11}\frac{d\lambda}{dN}=-\sqrt{6} \lambda^2(\Gamma-1)x\end{equation}

where $N=ln(a)$, $a$ is the scale factor. $\varsigma=1$ or $-1$ for
quintessence and phantom quintessence model. Here we should
emphasize that Eqs.(\ref{eqs9}-\ref{eqs11}) is not a dynamical
autonomous system since the parameter $\Gamma(\phi)$ is unknown. The
energy density fraction of dark energy scalar field is
\begin{equation}\label{eqs12}\Omega_{\phi}=\frac{\rho_{\phi}}{3M^2_{pl}H^2}=\varsigma x^2+y^2\end{equation}

while the equation of state of the dark energy scalar field is

\begin{equation}\label{eqs13}\gamma_{\phi}=1+w_{\phi}=1+\frac{\varsigma x^2-y^2}{\varsigma x^2+y^2}=\frac{2\varsigma x^2}{\varsigma x^2+y^2}=\frac{2\varsigma x^2}{\Omega_{\phi}}\end{equation}

\par In the other hand, it is more convenient to rewrite the dynamical system Eqs.(\ref{eqs9}-\ref{eqs10}) from the dependent variables $(x, y,
\lambda)$ directly to the observable quantities $(\Omega_{\phi},
\gamma_{\phi}, \lambda)$\cite{7}:

\begin{equation}\label{eqs14}\frac{d \Omega_{\phi}}{dN}=3(\gamma_b-\gamma_{\phi})\Omega_{\phi}(1-\Omega_{\phi})\end{equation}
\begin{equation}\label{eqs15}\frac{d \gamma_{\phi}}{dN}=(2-\gamma_{\phi})(\lambda \sqrt{3\varsigma \gamma_{\phi}\Omega_{\phi}}-3\gamma_{\phi})\end{equation}
\begin{equation}\label{eqs16}\frac{d\lambda}{dN}=-\sqrt{3} \lambda^2(\Gamma-1)\sqrt{\varsigma \gamma_{\phi}\Omega_{\phi}}\end{equation}

Eqs.(\ref{eqs14}-\ref{eqs16}) can reduce to quintessence when
$\varsigma=1$ (Eqs(3-5) in paper \cite{8} or Eqs.(17-18) in paper
\cite{7}) and reduce to phantom quintessence when $\varsigma=-1$
\cite{9}. Eqs.(\ref{eqs14}-\ref{eqs16}) are very useful to study the
cosmological implication of the evolutional behavior of the
dynamical system because the dynamical variables $\Omega_{\phi}$ and
$\gamma_{\phi}$ are the observable quantities. For example, $d
\gamma_{\phi}/dN>0$ corresponds to the thawing model and $d
\gamma_{\phi}/dN<0$ corresponds to the freezing model of the
evolution of equation of state of dark energy \cite{10,8}.

\par If $\Gamma=1$, Eq.(\ref{eqs11})(or Eqs.(\ref{eqs16})) will disappear, Eqs.(\ref{eqs9}-\ref{eqs10})(or
Eqs.(\ref{eqs14}-\ref{eqs15})) will become a two-dimensional
dynamical autonomous system. $\Gamma=1$ corresponds to the
exponential potential $V_0 e^{-\lambda \phi}$, which has been
studied in many literatures(e.g., see the references of paper
\cite{6}). However, the system described by
Eqs.(\ref{eqs9}-\ref{eqs11})(or Eqs.(\ref{eqs14}-\ref{eqs16})) are
not a dynamical autonomous system for other potentials because the
potential related parameter $\Gamma$ is unknown. Since $\lambda$ is
a function of quintessence scalar field $\phi$ and $\Gamma$ is also
a function of $\phi$, then $\Gamma$ can generally be expressed as a
function of $\lambda$. So if we consider $\Gamma$ as a function of
$\lambda$, namely $\Gamma=\Gamma(\lambda)$, then
Eqs.(\ref{eqs9}-\ref{eqs11})(or Eqs.(\ref{eqs14}-\ref{eqs16})) are
definitely a dynamical autonomous system, we therefore can study its
properties and dynamical evolution using the phase plane and
critical points analysis. Moreover, considering $\Gamma$ as a
function of $\lambda$ can cover many potentials beyond the
exponential potential\cite{6}.

\subsection{Dynamical System for tachyon and Phantom tachyon Scalar Field}
For the tachyon and phantom tachyon scalar field with lagrangian
$L=-V(\phi)\sqrt{1-\varsigma \dot \phi^2}$, we can define the
dimensionless variables as follows:

\begin{equation} \label{eqs17} x=\dot\phi, y=\frac{\kappa\sqrt{V}}{\sqrt3 H}, \lambda=\frac{V'}{\kappa V^{3/2}}, \Gamma=\frac{V V''}{V'^2} \end{equation}

Where $V'=dV(\phi)/d\phi, V''=d^2V(\phi)/d\phi^2 $.  Using
Eq.(\ref{eqs4}), Eq.(\ref{eqsadd1}),Eq.(\ref{eqs5})and
Eq.(\ref{eqs17}), the evolution of (phantom) tachyon can be
discribed in the following dynamical form \cite{12}:

\begin{equation}\label{eqs18} \frac{dx}{dN}=-\sqrt{3}(1-\varsigma x^2)(\sqrt{3}x+\varsigma \lambda y) \end{equation}
\begin{equation}\label{eqs19} \frac{dy}{dN}=\frac{\sqrt{3}}{2}y(\lambda xy+\frac{\sqrt{3} y^2(\varsigma x^2-\gamma_b)}{\sqrt{1-\varsigma x^2}}+\sqrt{3}\gamma_b) \end{equation}
\begin{equation}\label{eqs20} \frac{d\lambda}{dN}=\sqrt{3} x y\lambda^2(\Gamma-\frac{3}{2})\end{equation}

\par The density parameter of tachyon field $\Omega_{\phi}$, the equation of state $w_{\phi}$ are:

\begin{equation}\label{eqs21} \Omega_{\phi}=\frac{y^2}{\sqrt{1-\varsigma x^2}} \end{equation}
\begin{equation}\label{eqs22} \gamma_{\phi}=1+w_{\phi}=\varsigma x^2 \end{equation}

We can also rewrite the the dynamical system
Eqs.(\ref{eqs18}-\ref{eqs20}) from the dependent variables $(x, y,
\lambda)$ directly to the observable quantities $(\Omega_{\phi},
\gamma_{\phi}, \lambda)$:

\begin{equation}\label{eqs23}\frac{d \Omega_{\phi}}{dN}=3(\gamma_b-\gamma_{\phi})\Omega_{\phi}(1-\Omega_{\phi})\end{equation}
\begin{equation}\label{eqs24}\frac{d \gamma_{\phi}}{dN}=-2(1-\gamma_{\phi})[3\gamma_{\phi}+\lambda \sqrt{3\varsigma \gamma_{\phi} \Omega_{\phi}}(1-\gamma_{\phi})^{\frac{1}{4}}]\end{equation}
\begin{equation}\label{eqs25}\frac{d\lambda}{dN}=\sqrt{3\varsigma \gamma_{\phi}\Omega_{\phi}}(1-\gamma_{\phi})^{\frac{1}{4}}\lambda^2(\Gamma-\frac{3}{2})\end{equation}

The above equations Eqs.(\ref{eqs23}-\ref{eqs25}) are also obtained
in \cite{13, 13add}. The critical points and the dynamical evolution
of this system describing by Eqs.(\ref{eqs18}-\ref{eqs20}) or
Eqs.(\ref{eqs23}-\ref{eqs25}) had been studied in \cite{12}.

\par If $\Gamma=3/2$, Eq.(\ref{eqs20})(or
Eqs.(\ref{eqs25})) will disappear, Eqs.(\ref{eqs18}-\ref{eqs19})(or
Eqs.(\ref{eqs23}-\ref{eqs24})) will become a two-dimensional
dynamical autonomous system. For (phantom) tachyon scalar field,
$\Gamma=3/2$ corresponds to the case that the form of potential is
inverse square potential. But for other potentials, the system
described by Eqs.(\ref{eqs18}-\ref{eqs20})(or
Eqs.(\ref{eqs23}-\ref{eqs25})) will not be a dynamical autonomous
system any more since the potential related parameter $\Gamma$ is
unknown, and we can not exactly analyze the evolution of universe
like the inverse square potential any more. However, since $\lambda$
is the function of tachyon field $\phi$ and $\Gamma$ is also the
function of $\phi$, $\Gamma$ can be generally expressed as a
function of $\lambda$. So as the method used in\cite{6}, we can
consider $\Gamma$ as a function of $\lambda$,
$\Gamma=\Gamma(\lambda)$, then Eqs.(\ref{eqs18}-\ref{eqs20})(or
Eqs.(\ref{eqs23}-\ref{eqs25})) will become a dynamical autonomous
system, we therefore can study its properties and dynamical
evolution using the phase plane and critical points analysis. For
each form of function $\Gamma(\lambda)$, we can figure out the
detailed form of potential, so this method can cover many potentials
beyond the inverse square potential\cite{12}.
\par For the most simple case of potential $V(\phi)=V_0$, $\lambda=0$, Eqs.(\ref{eqs23}-\ref{eqs25}) become a very simple differential equation:

\begin{equation}\label{eqsadd13}\frac{d\Omega_{\phi}}{dN}=3(\gamma_b-\gamma_{\phi})\Omega_{\phi}(1-\Omega_{\phi}),~\frac{d\gamma_{\phi}}{dN}=-6\gamma_{\phi}(1-\gamma_{\phi})\end{equation}

We can get the exact solution for the above equations:

\begin{equation}\label{eqsadd14}\gamma_{\phi}=\frac{1}{1+c_1 e^{6N}}, ~\Omega_{\phi}=\frac{\sqrt{1+c_1 e^{6N}}}{\sqrt{1+c_1 e^{6N}}+c_2 (e^{6N})^{(1-\gamma_b)/2}}\end{equation}

Tachyon scalar field with this constant potential had been studied
in\cite{14}. According to the best-fit values of the parameters they
obtained, we can obtain the value of the integral constant
$c_1=0.0082$ and $c_2=0.59$ here. We know from Eq.(\ref{eqsadd14}),
when $N\rightarrow +\infty$, we have $\gamma_{\phi}\rightarrow 0$
and $\Omega_{\phi}\rightarrow 1$, the universe will be a de Sitter
like universe filled with the tachyon scalar field.

\subsection{Dynamical System for K-essence Scalar Field}
For the k-essence scalar field with lagrangian $L=-V(\phi)F(X)$, we
define the following dimensionless variables as the same as
Eq.(\ref{eqs17}):

\begin{equation}\label{eqs26} x=\dot\phi, \ \ y=\frac{\kappa\sqrt{V}}{\sqrt3 H}, \ \lambda=\frac{V'}{\kappa V^{3/2}}, \ \Gamma=\frac{V V''}{V'^2} \end{equation}

Using Eq.(\ref{eqs4}), Eq.(\ref{eqsadd1}),Eq.(\ref{eqs5})and
Eq.(\ref{eqs26}), we can get following dynamical system:

\begin{equation}\label{eqs27} \frac{dx}{dN}=\frac{-\sqrt{3}}{F_X+2XF_{XX}}[\sqrt{3}xF_X-y\lambda(F-2X F_X)]\end{equation}
\begin{equation}\label{eqs28} \frac{dy}{dN}=\frac{y}{2}[\sqrt{3}\lambda x y+6XF_X(\gamma_b-1)y^2-3\gamma_b F y^2+3\gamma_b] \end{equation}
\begin{equation}\label{eqs29}\frac{d\lambda}{dN}=\sqrt{3} x y\lambda^2(\Gamma-\frac{3}{2})\end{equation}

\par The density parameter of tachyon scalar field $\Omega_{\phi}$, the equation of state $w_{\phi}$ are:

\begin{equation}\label{eqs33} \Omega_{\phi}=(F-2XF_X)y^2=(F-xF')y^2 \end{equation}
\begin{equation}\label{eqs34} \gamma_{\phi}=\frac{xF'}{xF'-F}=-\frac{xy^2F'}{\Omega_{\phi}}\end{equation}

Where $F'=d F(X)/d x$ and $F''=d^2 F(X)/dx^2$, they both are the
functions of $x$. Using Eq.(\ref{eqs33}) and Eq.(\ref{eqs34}), we
can also rewrite the dynamical system Eq.(\ref{eqs27}),
Eq.(\ref{eqs28}) and Eq.(\ref{eqs29}) from the dependent variables
$(x, y, \lambda)$ directly to the observable quantities
$(\Omega_{\phi}, \gamma_{\phi}, \lambda)$:

\begin{equation}\label{eqs35}\frac{d \Omega_{\phi}}{dN}=3(\gamma_b-\gamma_{\phi})\Omega_{\phi}(1-\Omega_{\phi})\end{equation}
\begin{equation}\label{eqs36}\frac{d \gamma_{\phi}}{dN}=-\frac{\sqrt{3}(\lambda x y+\sqrt{3}\gamma_{\phi})[xF''(1-\gamma_{\phi})+F']}{xF''}=\sqrt{3}(\lambda x y+\sqrt{3}\gamma_{\phi})(\gamma_{\phi}-1-\frac{1}{2\Xi+1})\end{equation}
\begin{equation}\label{eqs37}\frac{d\lambda}{dN}=\sqrt{3}x y \lambda^2(\Gamma-\frac{3}{2})\end{equation}

where $\Xi=XF_{XX}/F_X=(xF''-F')/2F'$. We have pointed out the
relationship between quintessence and K-essence in previous section.
k-essence model described by Eq.(\ref{eqs7}) with a linear kinetic
function $F(X)=X+1$ can reduce to any quintessence model described
by Eq.(\ref{eqs5})\cite{4}. So all quintessence model with any
potentials can be considered as the special cases of k-essence
model. Moreover, it proved that the correspondence of the
exponential potential $V(\phi)=V_0 e^{-\lambda\phi}$ in quintessence
model is exactly the inverse square potential
$V(\phi)=(\frac{1}{2}\kappa \lambda \phi-c_1)^{-2}$ in k-essence
model. This can explain why the dynamical system
Eqs.(\ref{eqs9}-\ref{eqs11})(or Eqs.(\ref{eqs14}-\ref{eqs16})) of
quintessence reduces to two-dimensional autonomous system for
exponential potential($\Gamma=1$) while the same situation happens
for inverse square potential ($\Gamma=3/2$) in tachyon and k-essence
model. Author had considered another non-canonical scalar field
lagrangian defined as $L_{\phi}=Vf(B)$ with
$B=X/V$\cite{Tamanini2014}. It also includes the canonical
quintessence if we chose $f(B)=B-1$.

\par Let us discuss whether Eqs.(\ref{eqs35}-\ref{eqs37}) could be considered as an autonomous system.
Firstly, we realized that the variables $x$ and $y$ still appear in
Eq.(\ref{eqs36}) because we do not know the detailed form of
function $F(X)$ and can not figure out the solution of $x$ and $y$.
However, from Eqs.(\ref{eqs33}-\ref{eqs34}), we are sure that
variables $x$ and $y$ are the functions of variables $\Omega_{\phi}$
and $\gamma_{\phi}$. Since $\Xi(=XF_{XX}/F_X=(xF''-F')/2F')$ is a
function of $x$, it is also a function of $\Omega_{\phi}$ and
$\gamma_{\phi}$. Secondly for the potential related parameter
$\Gamma$, if we consider $\Gamma$ as a function of $\lambda$ just
similar with tachyon scalar field in subsection 3.2,
Eqs.(\ref{eqs35}-\ref{eqs37}) can eventually become a
three-dimensional dynamical autonomous system for any k-essence
models, and then we can easily study the critical points and the
dynamical evolution beyond the inverse square potential.

\par We can take tachyon model as a very simple example of k-essence model.
For (phantom) tachyon scalar field described in subsection 3.2, the
function $F(X)$ should take the form of $\sqrt{1-2\varsigma
X}=\sqrt{1-\varsigma x^2}$, and then Eqs.(\ref{eqs33}-\ref{eqs34})
will become Eqs.(\ref{eqs21}-\ref{eqs22}). We can obtain the
relationship as follows:

\begin{equation}\label{eqsadd12}x=\sqrt{\varsigma \gamma_{\phi}},~y=\sqrt{\Omega_{\phi}}(1-\gamma_{\phi})^{\frac{1}{4}},1+\frac{1}{2\Xi+1}=2-\gamma_{\phi} \end{equation}

Putting Eq.(\ref{eqsadd12}) into Eqs.(\ref{eqs35}-\ref{eqs37}),
these equations will reduce to Eqs.(\ref{eqs23}-\ref{eqs25}). That
means the dynamical system Eqs.(\ref{eqs35}-\ref{eqs37}) we derived
for k-essence model are correct.

\par Another example is lagrangian $p(\phi,
X)=f(\phi)(-X+X^2)$ which is proposed as a kinetically driven
quintessence\cite{16}. The pressure and energy density are given by

\begin{equation}\label{eqsadd5}p(\phi, X)=f(\phi)(-X+X^2);~~\rho=2X\frac{\partial p}{\partial X}-p=f(\phi)(-X+3X^2)\end{equation}

 Using Eqs.(\ref{eqs33}-\ref{eqs34}) and Eq.(\ref{eqsadd5}), we
can get following equation from Eqs.(\ref{eqs35}-\ref{eqs37}) :

\begin{equation}\label{eqsadd6}\frac{d \Omega_{\phi}}{dN}=3(\gamma_b-\gamma_{\phi})\Omega_{\phi}(1-\Omega_{\phi})\end{equation}
\begin{equation}\label{eqsadd7}\frac{d \gamma_{\phi}}{dN}=\frac{(\lambda\sqrt{3(4-3\gamma_{\phi})\Omega_{\phi}}+3\gamma_{\phi})(\gamma_{\phi}-2)(3\gamma_{\phi}-4)}{3\gamma_{\phi}-8}\end{equation}
\begin{equation}\label{eqsadd8}\frac{d\lambda}{dN}=\lambda^2\sqrt{3(4-3\gamma_{\phi})\Omega_{\phi}}(\Gamma-\frac{3}{2})\end{equation}

Above Eqs.(\ref{eqsadd6}-\ref{eqsadd8}) completely describe the
dynamical evolution of the kinetically driven quintessence.
Eq.(\ref{eqsadd8}) will vanish when $\Gamma=3/2$, then
Eqs.(\ref{eqsadd6}-\ref{eqsadd7}) is a two-dimensional autonomous
system which describes the dynamical evolution of the kinetically
driven quintessence with the lagrangian $p(\phi, X)=f(\phi)(-X+X^2)$
and potential $f(\phi)=(\frac{1}{2}\kappa \lambda \phi-c_1)^{-2}$.
In\cite{17}, authors obtained the two-dimensional dynamical
autonomous system with the dimensionless variables $(x, y)$ for this
type of k-essence model. They studied the phase-space properties and
the cosmological implications of the critical points in detail.
However, here we give the two-dimensional autonomous system
Eqs.(\ref{eqsadd6}-\ref{eqsadd7}) with the observational quantities
 $(\Omega_{\phi}, \gamma_{\phi})$ instead of the variables $(x, y)$. We can obtain the critical points of the
observational quantities $(\Omega_{\phi}, \gamma_{\phi})$ directly,
so it will be more convenient to study the properties of the
critical points and their cosmological implication with these
observational quantities. Furthermore, If we consider $\Gamma$ as a
function of $\lambda$, we can study the critical points and the
evolution of the universe beyond the inverse square potential, just
like the method used in\cite{12}.

The equation of state $w_{\phi}$$(=\gamma_{\phi}-1)$ and the
effective sound speed of perturbation $c_s^2$ can be obtained from
Eq.(\ref{eqsadd5}):

\begin{equation}\label{eqsadd9}w_{\phi}=\frac{X-1}{3X-1},~~~c_s^2=\frac{p_{,X}}{\rho_{,X}}=\frac{2X-1}{6X-1}=\frac{-(w_{\phi}+1)}{3(w_{\phi}+1)-8}=\frac{-\gamma_{\phi}}{3\gamma_{\phi}-8}\end{equation}

\begin{figure}
\begin{minipage}[t]{0.45\linewidth}
\centering
\includegraphics[scale=0.38,origin=c,angle=0]{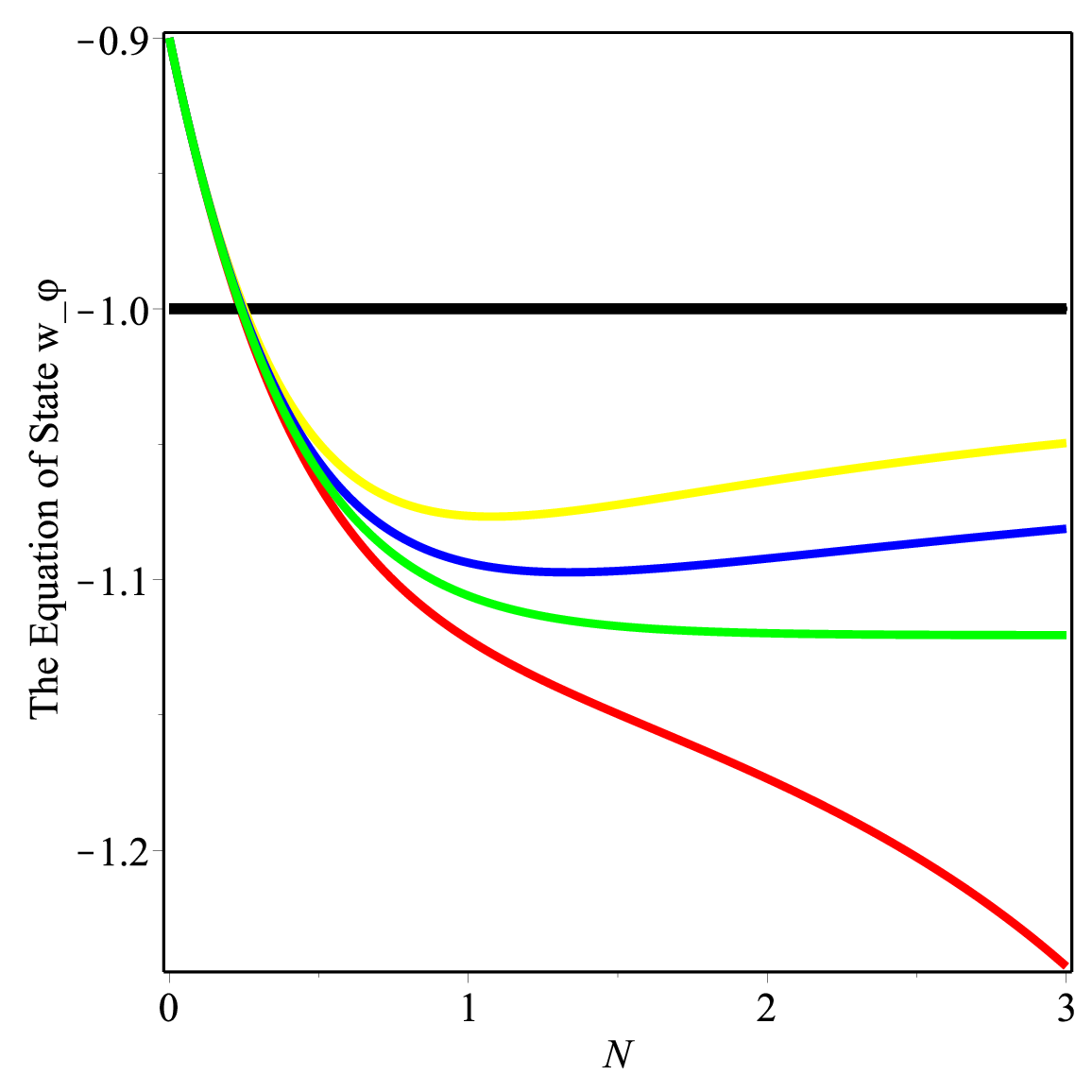}
 \caption{The evolution of the equation of state $w_{\phi}$ with respect to $N$ when
$w_{\phi}$ cross the Phantom Line. Fig.\ref{fig7} and Fig.\ref{fig8}
are plotted using differential equations
(\ref{eqsadd6}-\ref{eqsadd8}).} \label{fig7}
\end{minipage}
\hfill
\begin{minipage}[t]{0.50\linewidth}
\centering
\includegraphics[scale=0.38,origin=c,angle=0]{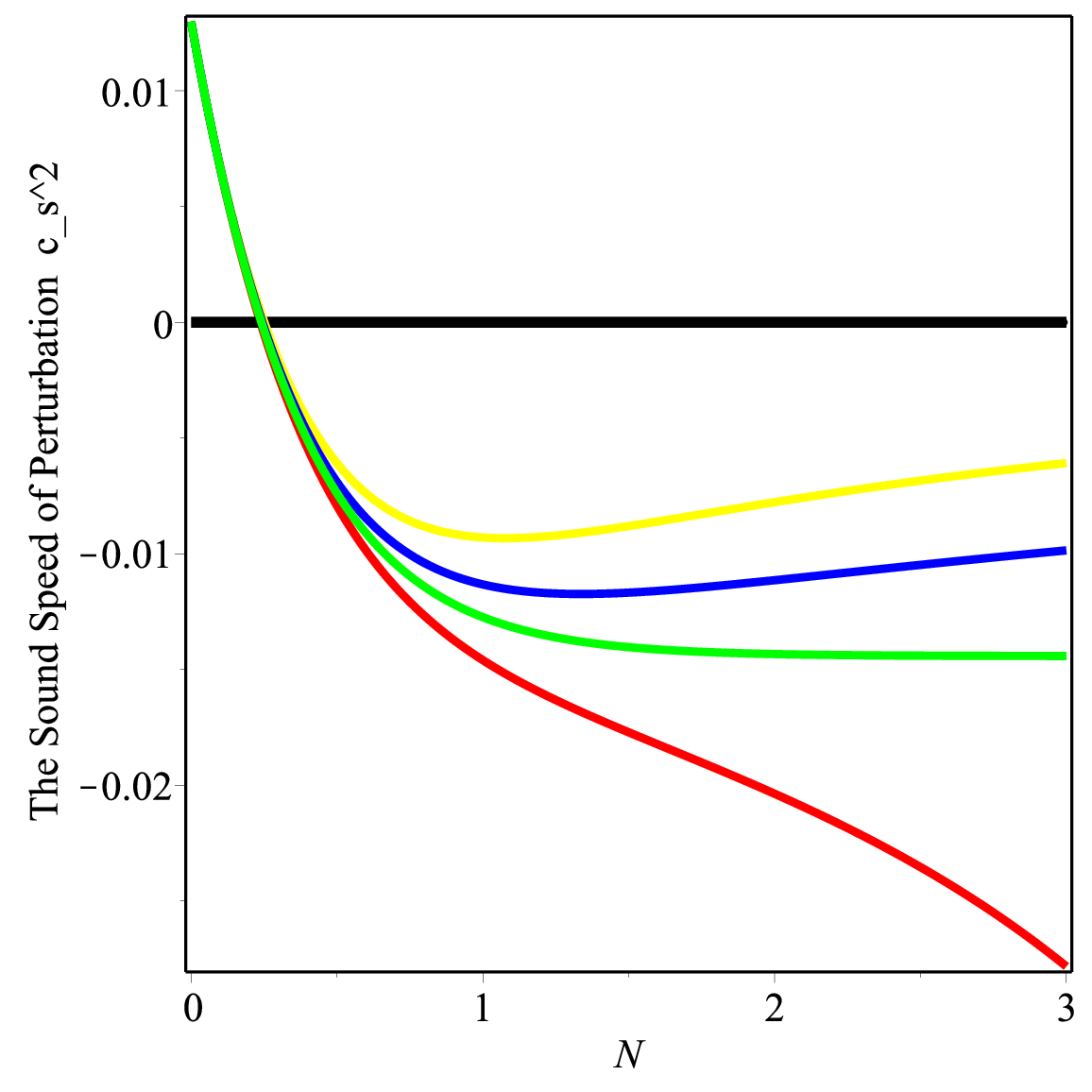}
\caption{The evolution of $c_s^2$ with respect to $N$. All the
initial conditions are the same as Fig.\ref{fig7}. Fig.\ref{fig7}
and Fig.\ref{fig8} are plotted under the model of kinetically driven
quintessence$^{\spadesuit}$.}\label{fig8}
\end{minipage}
\end{figure}

\begin{figure}
\begin{minipage}[t]{0.47\linewidth}
\centering
\includegraphics[scale=0.38,origin=c,angle=0]{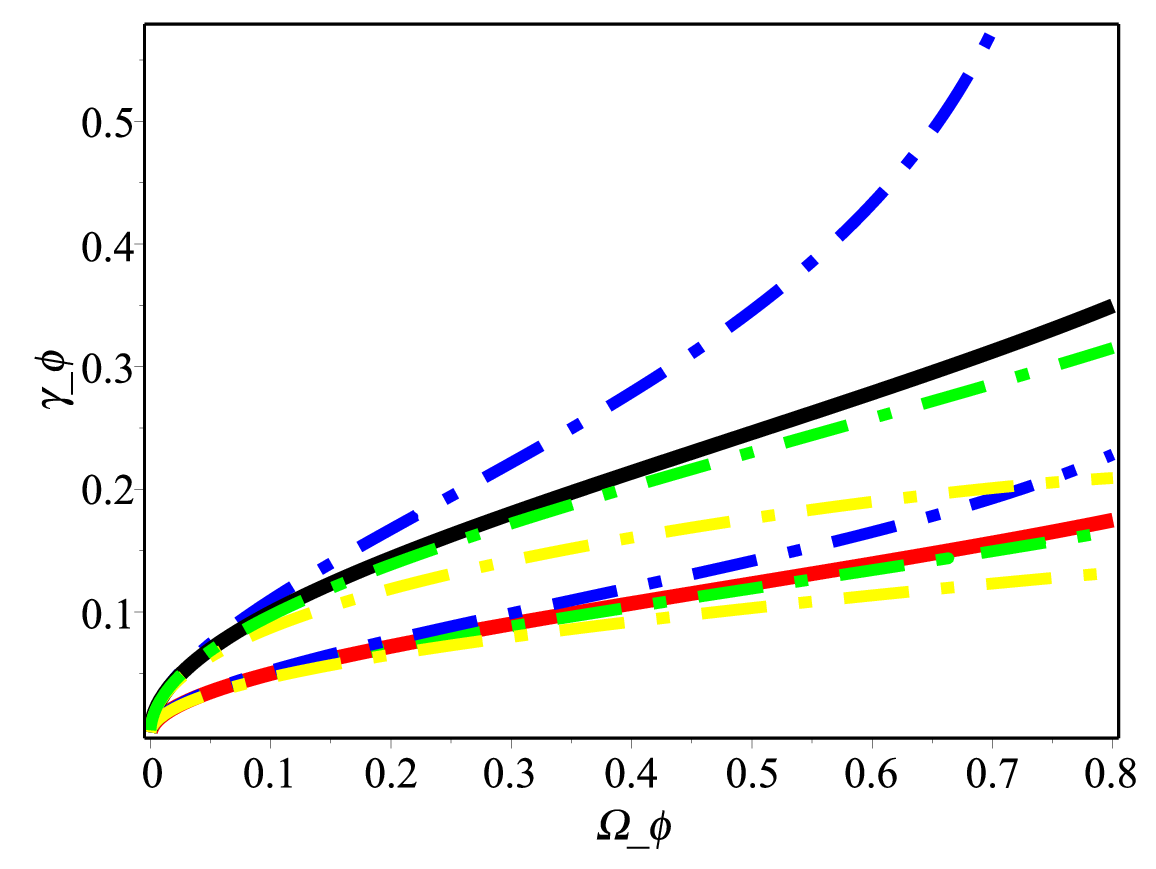}
 \caption{The evolutional behavior of $\gamma_{\phi}$ with respect to
 $\Omega_{\phi}$. $\lambda_0=-0.2, -0.4$ for red and black
 solid curve respectively. The initial condition when $\Omega_{\phi}=0$ is
($\gamma_{\phi}=0$, $\lambda=-0.2$) and ($\gamma_{\phi}=0$,
$\lambda=-0.4$) for the three dashed blue, green and yellow curves
around the red and black solid line respectively.} \label{fig5}
\end{minipage}
\hfill
\begin{minipage}[t]{0.47\linewidth}
\centering
\includegraphics[scale=0.38,origin=c,angle=0]{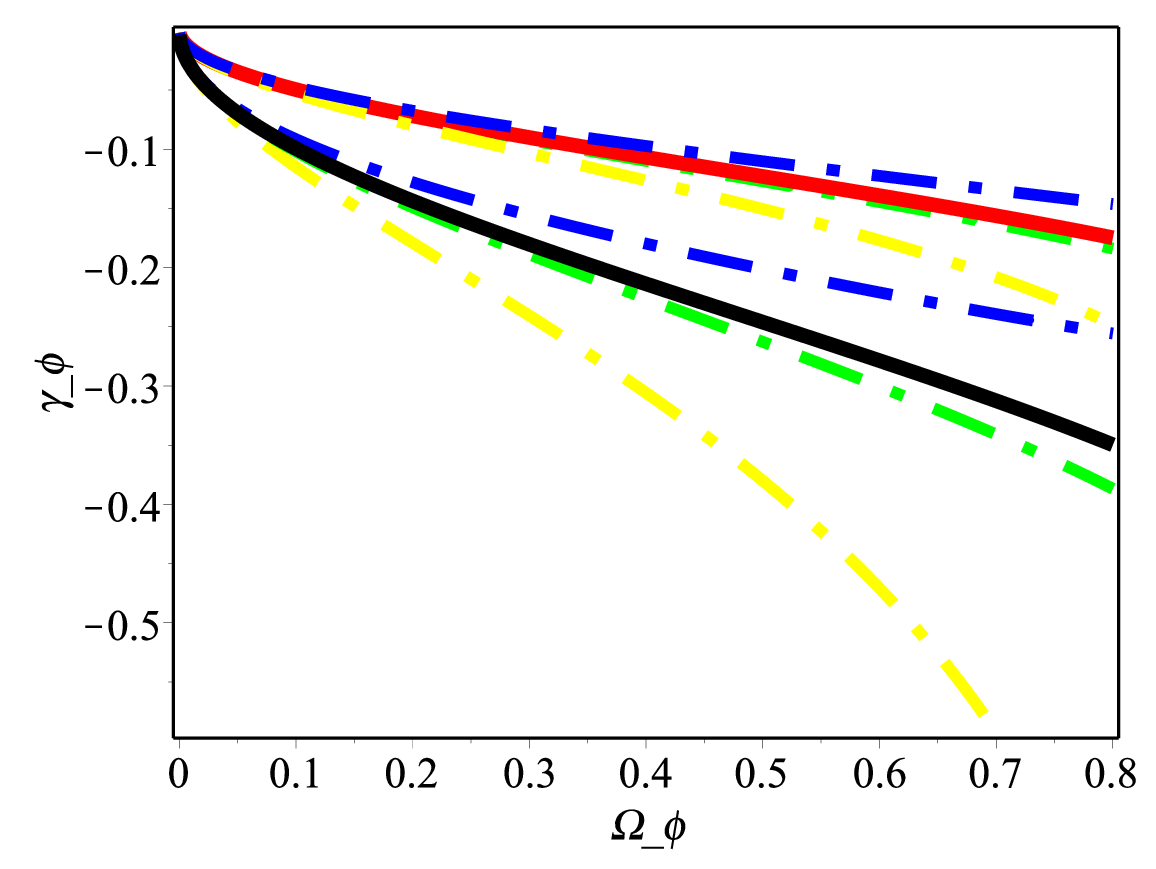}
 \caption{The evolutional behavior of $\gamma_{\phi}$ with respect to
 $\Omega_{\phi}$. All the initial conditions are the same as Fig.\ref{fig5} except $\lambda_0=0.2,
0.4$. Solid curve in Fig.\ref{fig5} and Fig.\ref{fig6} are plotted
using Eq.(\ref{eqsadd11}), all other curves plotted using
differential equations (\ref{eqsadd6}-\ref{eqsadd8}) with different
potentials(namely, $\Gamma=1, 3/2, 2$).}\label{fig6}
\end{minipage}
\end{figure}

It is interesting that $w_{\phi}$ could be larger or less than $-1$
for this kind of k-essence model(Fig.\ref{fig7}). $w_{\phi}$ is
larger than $-1$ when $X<1/3$ (Noted $X\geq 0$ since
$X=\dot\phi^2/2$) or $X>1/2$, and less than $-1$ when $1/2>X>1/3$.
So it is possible for the equation of state $w_{\phi}$ crossing the
Phantom Line. We have solved numerically
Eqs.(\ref{eqsadd6}-\ref{eqsadd8}) for several different potentials
and plot the evolution of $w_{\phi}$ crossing the Phantom Line in
Fig.\ref{fig7}. However, when the equation of state $w_{\phi}$
crosses the Phantom Line, the effective sound speed of perturbations
$c_s^2$ will change its sign from positive to negative
simultaneously(Fig.\ref{fig8}). For the stability with respect to
the general metric and matter perturbation, the condition
$c_s^2\geq0$ is necessary, so the background models with $c_s^2<0$
are violently unstable and do not have any physical significance.
Therefore this model of transition is not realistic \cite{18,19}.
 However, we show point out that, not all the
kinetically driven quintessence we discussed here will cross the
phantom divide in the future(see the Fig.\ref{fig5} and
Fig.\ref{fig6}, and the discussion about them). It is determined by
the potentials and initial conditions. Here we just show the
possibility that the kinetically driven quintessence can cross the
phantom divide.

\footnotetext{$\spadesuit$ The curves in yellow, blue, green and red
in Fig.\ref{fig7} and Fig.\ref{fig8} are for $\Gamma=0, ~1, ~3/2,
~2$, corresponding to the potentials being the form of $V_0(\phi+c),
~V_0 e^{c\phi}, ~V_0(\phi+c)^{-2}, ~V_0(\phi+c)^{-1}$ respectively.
The initial conditions for these four potentials are the same,
$\gamma_{\phi 0}=0.1,~\Omega_{\phi 0}=0.7$ and $\lambda=0.1$ when
$N=0$(at present time).}

If the equation of state $w_{\phi}$ is near $-1$(so $\gamma_{\phi}
\sim 0$)  for the kinetically driven
 quintessence model, we can drop terms of higher order
 in $\gamma_{\phi}$, and further get a simple differential equation for $\gamma_{\phi}$ with the dependent variable from $N$ to
$\Omega_{\phi}$ from Eqs.(\ref{eqsadd6}-\ref{eqsadd7}):

\begin{equation}\label{eqsadd10}\frac{d \gamma_{\phi}}{d \Omega_{\phi}}=-\frac{2\lambda}{\sqrt{3\Omega_{\phi}}(1-\Omega_{\phi})}-\frac{\gamma_{\phi}}{\Omega_{\phi}(1-\Omega_{\phi})}\end{equation}

 We assume that $\lambda$ is approximately constant($=\lambda_0$) when $\gamma_{\phi}$ is near $0$, so
 that the above equation can be solved exactly:

\begin{equation}\label{eqsadd11}\gamma_{\phi}=1+w_{\phi}=-\frac{\sqrt{3}}{3}\lambda_0\frac{1-\Omega_{\phi}}{\Omega_{\phi}}\left(ln\left[\frac{1-\sqrt{\Omega_{\phi}}}{1+\sqrt{\Omega_{\phi}}}\right]+\frac{2\sqrt{\Omega_{\phi}}}{1-\Omega_{\phi}}\right)\end{equation}

we have chosen the boundary condition that $\gamma_{\phi}=0$ at
$\Omega_{\phi}=0$. Eq.(\ref{eqsadd11}) gives the general behaviors
of all the kinetic driven quintessence model for all sufficiently
flat potentials.

Fig.\ref{fig5} and Fig.\ref{fig6} show how accurate the analytic
result(Eq.(\ref{eqsadd11})) is and the tiny differences among
different potentials. solid black and red curves are plotted using
Eq.(\ref{eqsadd11}) directly, and all other curves are numerically
plotted from Eqs.(\ref{eqsadd6}-\ref{eqsadd8}).

The solid curves(red and black) in Fig.\ref{fig5} and Fig.\ref{fig6}
show the general relationship(Eq.(\ref{eqsadd11})) of
$\gamma_{\phi}$, $\Omega_{\phi}$ and $\lambda_0$. Around each solid
curve are the dashed color curves(blue, green and yellow), which are
the numerical results of Eqs.(\ref{eqsadd6}-\ref{eqsadd8}) with
different values of $\Gamma$ since different values of $\Gamma$
corresponds to different form of potentials: dashed blue curve for
$\Gamma=1$($~V_0 e^{c\phi}$),
 dashed green curve for $\Gamma=3/2$($~V_0(\phi+c)^{-2}$) and dashed yellow curve
for $\Gamma=2$($V_0(\phi+c)^{-1}$). For the dashed green curve,
$\Gamma=3/2$ and then $\lambda$ is a constant, so it is easily
understood that why the difference between dashed green curve and
red(black) solid curve is so small. It also demonstrated that
Eq.(\ref{eqsadd11}) is valid when the value of $\gamma_{\phi}$ is
small.  The only difference between Fig.\ref{fig5} and
Fig.\ref{fig6} is the opposite values of $\lambda_0$. The initial
values of $\lambda$ when $\Omega_{\phi}=0$ are $-0.2$(red) and
$-0.4$(black) in Fig.\ref{fig5} while $0.2$(red) and $0.4$(black) in
Fig.\ref{fig6}. We find that the red solid curve and the dashed
color curves around it in both Fig.\ref{fig5} and Fig.\ref{fig6} are
more close to the Phantom Line($w_{\phi}=-1$). So the smaller the
initial value of $|\lambda|$ is(i.e., the more flat the potential
is), the less deviation the equation of state $w_{\phi}$ is from
$-1$. It is interesting that, for different potentials $~V_0
e^{c\phi}$, $V_0(\phi+c)^{-2}$ and $V_0(\phi+c)^{-1}$, the state of
equation $w_{\phi}$ can larger or less than $-1$, and also can
increase or decrease with respect to $\Omega_{\phi}$, only depending
on the initial value of $\lambda$(determined by the initial value of
$\phi$). Unlike the evolution in Fig.\ref{fig7} and Fig.\ref{fig8},
$w_{\phi}$ plotted in Fig.\ref{fig5} and Fig.\ref{fig6} does not
cross Phantom Line due to the different choice of the initial
conditions.

\subsection{Dynamical System for general non-canonical scalar
field}

In the last subsection of section 3, we consider the general non-
canonical scalar field with a lagrangian $L=F(X)-V(\phi)$ filled in
a spatially flat Fridmann-Lemaitre-Robertson-Walker(FLRW) cosmology.
The motion of this scalar field and the evolution of the universe
are described by Eqs.(\ref{eqs4}-\ref{eqsadd1}) and Eq.(\ref{eqs8}).

\par In order to obtain the autonomous system we define the
variables as follows\cite{20}$^{\dag}$:

\footnotetext{$\dag$ We think the definition of variables in
\cite{20} can not work if $2XF_X-F<0$. Take a very simple example of
phantom quintessence with $F=-X$, $2XF_X-F=-X<0$, then the
definition of variables in \cite{20} is undefined. Therefore, we
introduce the parameter $\varsigma$, $\varsigma=1$ for $2XF_X-F>0$
and $\varsigma=-1$ for $2XF_X-F<0$.}

\begin{equation}\label{eqs38} x=\frac{\sqrt{(2XF_X-F)\varsigma}}{\sqrt{3}M_{pl}H},y=\frac{\sqrt{V}}{\sqrt{3}M_{pl}H},\sigma=-\frac{M_{pl}V_{\phi}}{V}\sqrt{\frac{2X}{3(2XF_X-F)\varsigma}}sign(\dot{\phi})\end{equation}

Using Eq.(\ref{eqs8})and Eq.(\ref{eqs38}),
Eq.(\ref{eqs4}-\ref{eqsadd1}) can be expressed as following
dynamical system:

\begin{equation}\label{eqs39}\frac{dx}{dN}=\frac{3}{2}[\sigma \varsigma y^2-x(w_k+1)]+\frac{3}{2}x[(1+w_b)(1-y^2)+(w_k-w_b)\varsigma x^2]\end{equation}
\begin{equation}\label{eqs40}\frac{dy}{dN}=-\frac{3}{2}\sigma xy+\frac{3}{2} y[(1+w_b)(1-y^2)+(w_k-w_b)\varsigma x^2]\end{equation}
\begin{equation}\label{eqs41}\frac{d\sigma}{dN}=-3 \sigma^2 x(\Gamma-1)+\frac{3\sigma[2\Xi(w_k+1)+w_k-1]}{2(2\Xi+1)(w_k+1)}(w_k+1-\frac{\sigma y^2}{\varsigma x})\end{equation}

where

\begin{equation}\label{eqs42}w_k=\gamma_k-1=\frac{F}{2XF_X-F},~w_{\phi}=\frac{p_{\phi}}{\rho_{\phi}}=\frac{w_k\varsigma x^2-y^2}{\varsigma x^2+y^2},~\Omega_{\phi}=\varsigma x^2+y^2,~\Xi=XF_{XX}/F_X\end{equation}

\par From Eq.(\ref{eqs42}), we can get following relationships:

\begin{equation}\label{eqs43} x^2=\frac{\gamma_{\phi}\varsigma}{\gamma_k}\Omega_{\phi},~~~~y^2=\frac{\gamma_k-\gamma_{\phi}}{\gamma_k}\Omega_{\phi}\end{equation}

 Using Eqs.(\ref{eqs42}-\ref{eqs43}), the dynamical system Eqs.(\ref{eqs39}-\ref{eqs41}) can be rewritten with the variables
 $(\Omega_{\phi}, \gamma_{\phi}, \sigma)$ related with observable quantities:

\begin{equation}\label{eqs44}\frac{d \Omega_{\phi}}{dN}=3(\gamma_b-\gamma_{\phi})\Omega_{\phi}(1-\Omega_{\phi})\end{equation}
\begin{equation}\label{eqs45}\frac{d \gamma_{\phi}}{dN}=3\left[\varsigma\gamma_{\phi}(\gamma_{\phi}-1-\frac{1}{2\Xi+1})+2\sigma\sqrt{\frac{\varsigma\Omega_{\phi}\gamma_{\phi}}{\gamma_k}}\frac{(\gamma_k-\gamma_{\phi})(\Xi+1)}{(2\Xi+1)\gamma_k}\right]\end{equation}
\begin{equation}\label{eqs46}\frac{d\sigma}{dN}=-3\sigma\left(\sigma (\Gamma-1)\sqrt{\frac{\varsigma\Omega_{\phi}\gamma_{\phi}}{\gamma_k}}+(\frac{1}{2\Xi+1}-\frac{\gamma_k}{2})\left[\sigma \sqrt{\frac{\varsigma\Omega_{\phi}\gamma_{\phi}}{\gamma_k}}(\frac{1}{\gamma_k}-\frac{1}{\gamma_{\phi}})+1\right]\right)\end{equation}

 Beside the three variables $(\Omega_{\phi}, \gamma_{\phi}, \sigma)$, there are other quantities in Eqs.(\ref{eqs45}-\ref{eqs46}) which are not constant:
  $\Xi, \gamma_k$ and $\Gamma$. We know that $\gamma_{\phi}=p_{\phi}/\rho_{\phi}+1=\gamma_{\phi}(X,\phi)=\gamma_{\phi}(\dot\phi, \phi)$ and
 $\sigma=\sigma(\dot\phi, \phi)$. So generally speaking, we can obtain the expressions of $\phi=\phi(\gamma_{\phi}, \sigma)$ and
 $\dot \phi=\dot \phi(\gamma_{\phi}, \sigma)$. For the parameters $\Xi, \gamma_k$ and $\Gamma$, we know $\Xi=\Xi(\dot\phi), \gamma_k=\gamma_k(\dot \phi)$
 and $\Gamma=\Gamma(\phi)$, so they can be expressed as $\Xi(\gamma_{\phi}, \sigma), \gamma_k(\gamma_{\phi}, \sigma)$ and
 $\Gamma(\gamma_{\phi}, \sigma)$. Therefore, Eqs.(\ref{eqs44}-\ref{eqs46}) could in principal be a dynamical autonomous system though
 it is actually quite complicated.
\par We take two simple examples to check the correctness of Eqs.(\ref{eqs44}-\ref{eqs46}) and study its properties. First example is,
we know that general non-canonical scalar field with lagrangian
$F(X)-V(\phi)$ will reduce to (phantom) quintessence if
$F(X)=\varsigma X$, and Eqs.(\ref{eqs45}-\ref{eqs46}) should reduce
to Eqs.(\ref{eqs15}-\ref{eqs16})(Eq.(\ref{eqs44}) has the same form
with Eq.(\ref{eqs14})). $F(X)=\varsigma X$ makes $\Xi=0$,
$\gamma_k=2$ and $\sigma=\sqrt{\frac{2}{3}}\lambda$, then we found
Eqs.(\ref{eqs45}-\ref{eqs46}) have the same form with
Eqs.(\ref{eqs15}-\ref{eqs16}).

Another example is that, if the potential $V(\phi)=-1$ in K-essence
and the Potential $V(\phi)=0$ in general non-canonical scalar field,
these two scalar field models will have the same form of lagrangian
$L=F(X)$, then both will reduce to the so-called purely kinetic
united model\cite{21, 22}. In this case, $d\lambda/dN=0$ and
$d\sigma/dN=0$, then both Eqs.(\ref{eqs35}-\ref{eqs37}) and
Eqs.(\ref{eqs44}-\ref{eqs46}) will reduce to a two-dimensional
dynamical system as follows:

\begin{equation}\label{eqs51}\frac{d \Omega_{\phi}}{dN}=3(\gamma_b-\gamma_{\phi})\Omega_{\phi}(1-\Omega_{\phi})\end{equation}
\begin{equation}\label{eqs52}\frac{d \gamma_{\phi}}{dN}=3\gamma_{\phi}(\gamma_{\phi}-1-\frac{1}{2\Xi+1})\end{equation}

For the case of purely kinetic united model $L=F(X)$, we know from
Eq.(\ref{eqs38}) and Eq.(\ref{eqs42}) that $\gamma_{\phi}$ is a
function of $X$, $\gamma_{\phi}=\gamma_k=\frac{F}{2XF_X-F}$. In the
meantime, $\Xi$ is also a function of $X$ because
$\Xi=XF_{XX}/F_X=(xF''-F')/2F'$. So generally speaking, $\Xi$ can be
expressed as a function of $\gamma_{\phi}$. Then the system of
Eqs.(\ref{eqs51}-\ref{eqs52}) could be a two-dimensional autonomous
dynamical system. For some special cases of $F(X)$, we can even get
the exact solution for $\Omega_{\phi}$ and $\gamma_{\phi}$. We take
two examples to illustrate our viewpoints.
\par The most simple is the case that $\Xi$ is a constant. We know
that $\Xi=\frac{2-\alpha}{2(\alpha-1)}=\frac{XF_{XX}}{F_X}$, we can
integrate and get the form for $F(X)$:

\begin{equation}\label{eqs54} F(X) =C X^{\frac{\alpha}{2(\alpha-1)}}+X_0 \end{equation}

 We can also get the exact solution for $\Omega_{\phi}$ and $\gamma_{\phi}$ from Eqs.(\ref{eqs51}-\ref{eqs52}):

\begin{equation}\label{eqs53}\gamma_{\phi}(N)=\frac{1}{c_3 e^{3\alpha N}+1/\alpha},~~ \Omega_{\phi}(N)=\frac{1}{c_4(c_3\alpha e^{3\alpha N}+1)^{-1}e^{3(\alpha-\gamma_b)N}+1}\end{equation}

 where $\alpha=1+\frac{1}{2\Xi+1}$, $C$, $X_0$, $c_3$ and $c_4$ are the integral constants. If we set the equation of state of dark energy $\gamma_{\phi}(0)=\gamma_0
 $ and energy density of dark energy $ \Omega_{\phi}(0)= \Omega_0$ at present $N=0$, we then get $c_3=\frac{\alpha-\gamma_0}{\alpha \gamma_0}$ and $c_4=
 \frac{\alpha(1-\Omega_0)}{\Omega_0\gamma_0}$.

\begin{figure}
\begin{minipage}[t]{0.48\linewidth}
\centering
\includegraphics[scale=0.38,origin=c,angle=0]{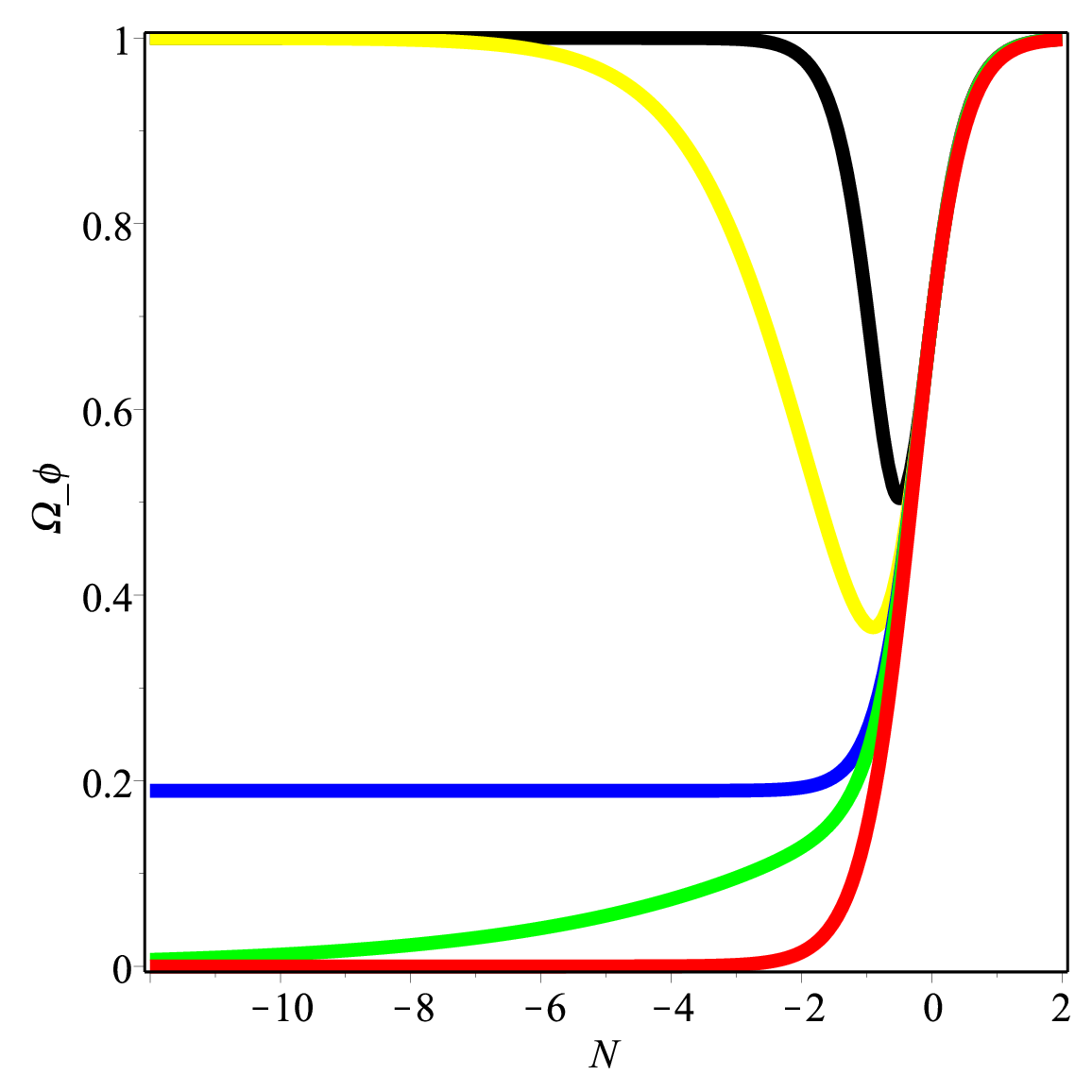}
 \caption{The evolution of $\Omega_{\phi}$ with respect to $N$
when $F(X) =C X^{\frac{\alpha}{2(\alpha-1)}}+X_0$ and $\gamma_b=1$,
initial condition is chosen as $\gamma_{\phi}=0.1$ and
$\Omega_{\phi}=0.7$ at present($N=0$). The color lines from the top
to the bottom are plotted with $\alpha=0.3, ~0.9, ~1.0, ~4/3,
~2.0$.}\label{fig2}
\end{minipage}
\hfill
\begin{minipage}[t]{0.46\linewidth}
\centering
\includegraphics[scale=0.38,origin=c,angle=0]{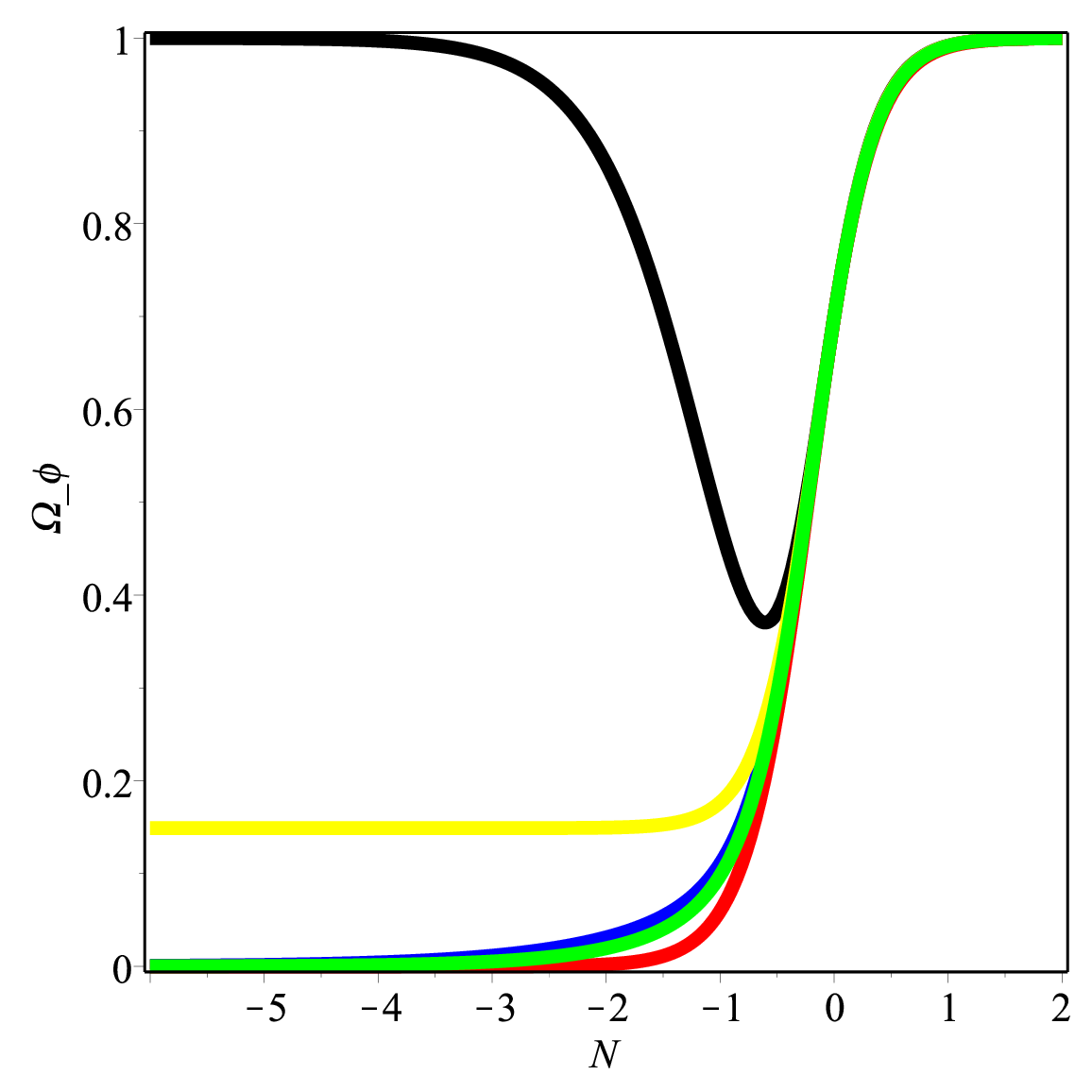}
 \caption{The evolution of $\Omega_{\phi}$ with respect to $N$
when $F(X) =C X^{\frac{\alpha}{2(\alpha-1)}}+X_0$. The initial
condition and the values of the parameter $\alpha$ are the same as
Fig.\ref{fig2} except $\gamma_b=4/3$. Fig.\ref{fig2} and
Fig.\ref{fig4} are plotted using Eq.(\ref{eqs53}).}\label{fig4}
\end{minipage}
\end{figure}

 For the cosmic evolution in very early time, $N$ is negative and $|N|$ is very large,
 we get $\gamma_{\phi}\approx \alpha$ from Eq.(\ref{eqs53}), and then energy density of scalar field will behave as $\rho_{\phi}\sim a^{-3\alpha}$.

 For the cosmic evolution in late time, $N$ is positive and very large, $\gamma_{\phi}\approx 0$ and $\Omega_{\phi}\approx 1$, scalar field behaves as
 the cosmological constant.

  Noted that there is only kinetic term in the lagrangian which gives the cosmological constant solution.
 Moreover,

 From Eq.(\ref{eqs53}) that $\Omega_{\phi}$ might not be 0 in the very early time. Its value depends on the value of $\alpha$ comparing with
 the value of $\gamma_b$($\gamma_b=1$ for matter and $\gamma_b=4/3$ for radiation).

 We have plotted the evolution of $\Omega_{\phi}$ with respect to $N$
 for different $\alpha$ and different $\gamma_b$ in Fig.\ref{fig2} and Fig.\ref{fig4}, and

 Fig.\ref{fig2} and Fig.\ref{fig4} show that the value of $\Omega_{\phi}$ in the very early
 time could be 1, 0 or a positive constant which is less then 1.

  When $N \rightarrow -\infty$,  we can get $\Omega_{\phi}(N) \approx 1/(c_4 e^{3(\alpha
 -\gamma_b)N}+1)$ from Eq.(\ref{eqs53}).

 Then $\Omega_{\phi}\rightarrow 1$ for $\alpha>\gamma_b$, $\Omega_{\phi}\rightarrow 1/(c_4+1)$ for $\alpha=\gamma
 _ b$ and $\Omega_{\phi}\rightarrow 0$ for $\alpha<\gamma_b$.

  If $\Omega_{\phi}>0$ in very early time, it would be very interesting to investigate its impact on the
 evolutional history of early universe.

The second case is the lagrangian as follows:

\begin{equation}\label{eqsadd15}F(X)=A_1\sqrt{X}-A_2 X^{\beta}\end{equation}

where $A_1$, $A_2$, and $\beta$ are constants. This form of $F(X)$
was proposed in \cite{21, 23}. From Eq.(\ref{eqs42}), we can get
that $\Xi=(2\beta-\gamma_{\phi})/2\gamma_{\phi}$, then dynamical
system Eqs.(\ref{eqs51}-\ref{eqs52}) becomes the following
equations:

\begin{equation}\label{eqsadd16}\frac{d \Omega_{\phi}}{dN}=3(\gamma_b-\gamma_{\phi})\Omega_{\phi}(1-\Omega_{\phi})\end{equation}
\begin{equation}\label{eqsadd17}\frac{d \gamma_{\phi}}{dN}=3\gamma_{\phi}(\frac{2\beta-1}{2\beta} \gamma_{\phi}-1)\end{equation}

Solving above differential equations, we obtain the following exact
solution for $\gamma_{\phi}$ and $\Omega_{\phi}$:

\begin{equation}\label{eqsadd18}\gamma_{\phi}(N)=\frac{1}{c_5 e^{3N}+1/\eta},~~ \Omega_{\phi}(N)=\frac{1}{c_6(c_5\eta e^{3N}+1)^{-\eta}e^{3(\eta-\gamma_b)N}+1}\end{equation}

 where $c_5$ and $c_6$ are the integral constants, $\eta=\frac{2\beta}{2\beta-1}$, is also a
 constant.  Eq.(\ref{eqsadd18}) is very similar but a little different with the evolution described in Eq.(\ref{eqs53}):

$$\gamma_{\phi}(N)=\frac{1}{c_3 e^{3\alpha N}+1/\alpha},~~\Omega_{\phi}(N)=\frac{1}{c_4(c_3\alpha e^{3\alpha N}+1)^{-1}e^{3(\alpha-\gamma_b)N}+1}$$

 \begin{figure}
\begin{minipage}[t]{0.48\linewidth}
\centering
\includegraphics[scale=0.38,origin=c,angle=0]{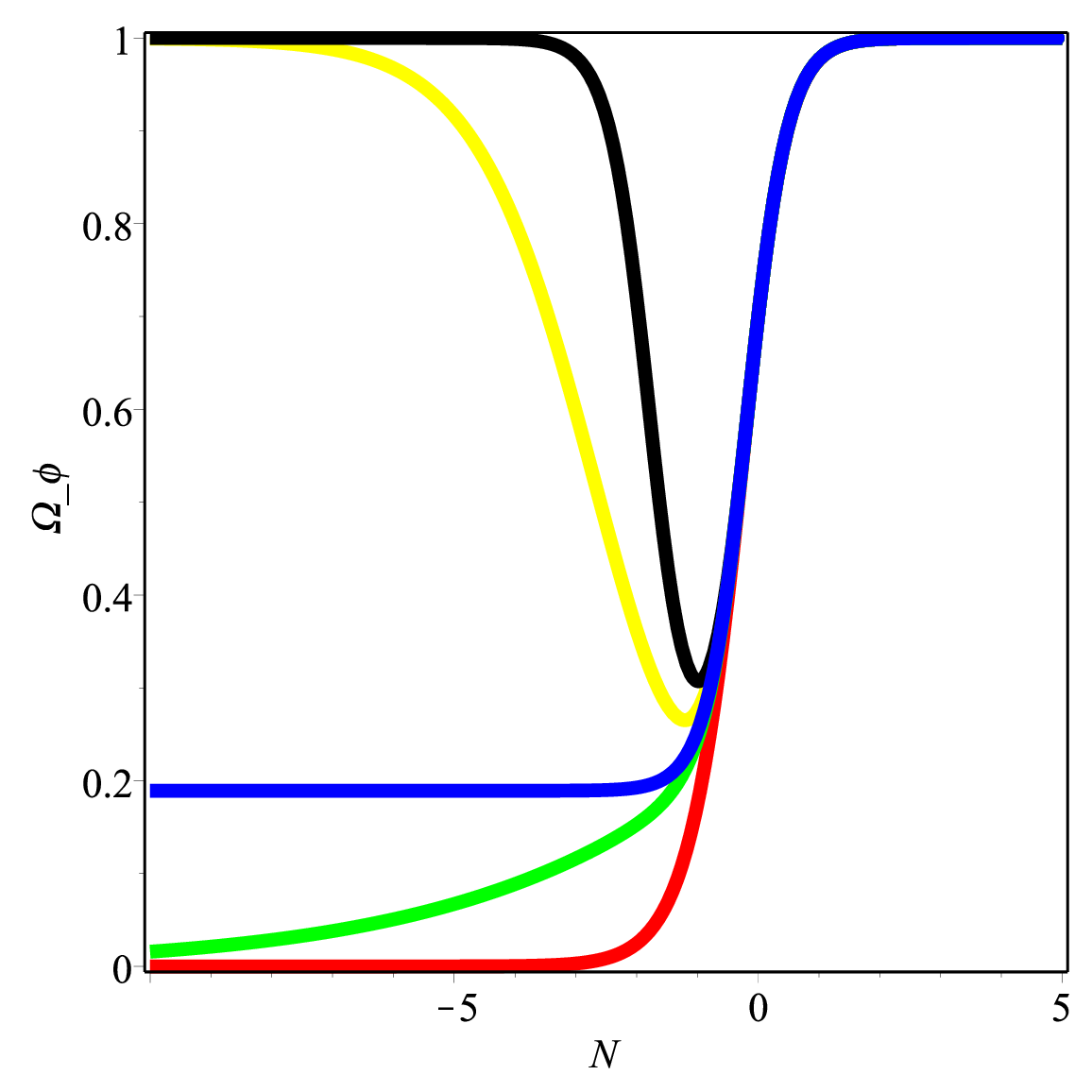}
 \caption{The evolution of $\Omega_{\phi}$ with respect to $N$
when $F(X)=A_1\sqrt{X}-A_2 X^{\beta}$ and $\gamma_b=1$, initial
condition is chosen as $\gamma_{\phi}=0.1$ and $\Omega_{\phi}=0.7$
at present($N=0$). The color lines from the top to the bottom are
plotted with $\eta=0.3, ~0.9, ~1.0, ~4/3, ~2.0$.}\label{fig9}
\end{minipage}
\hfill
\begin{minipage}[t]{0.46\linewidth}
\centering
\includegraphics[scale=0.38,origin=c,angle=0]{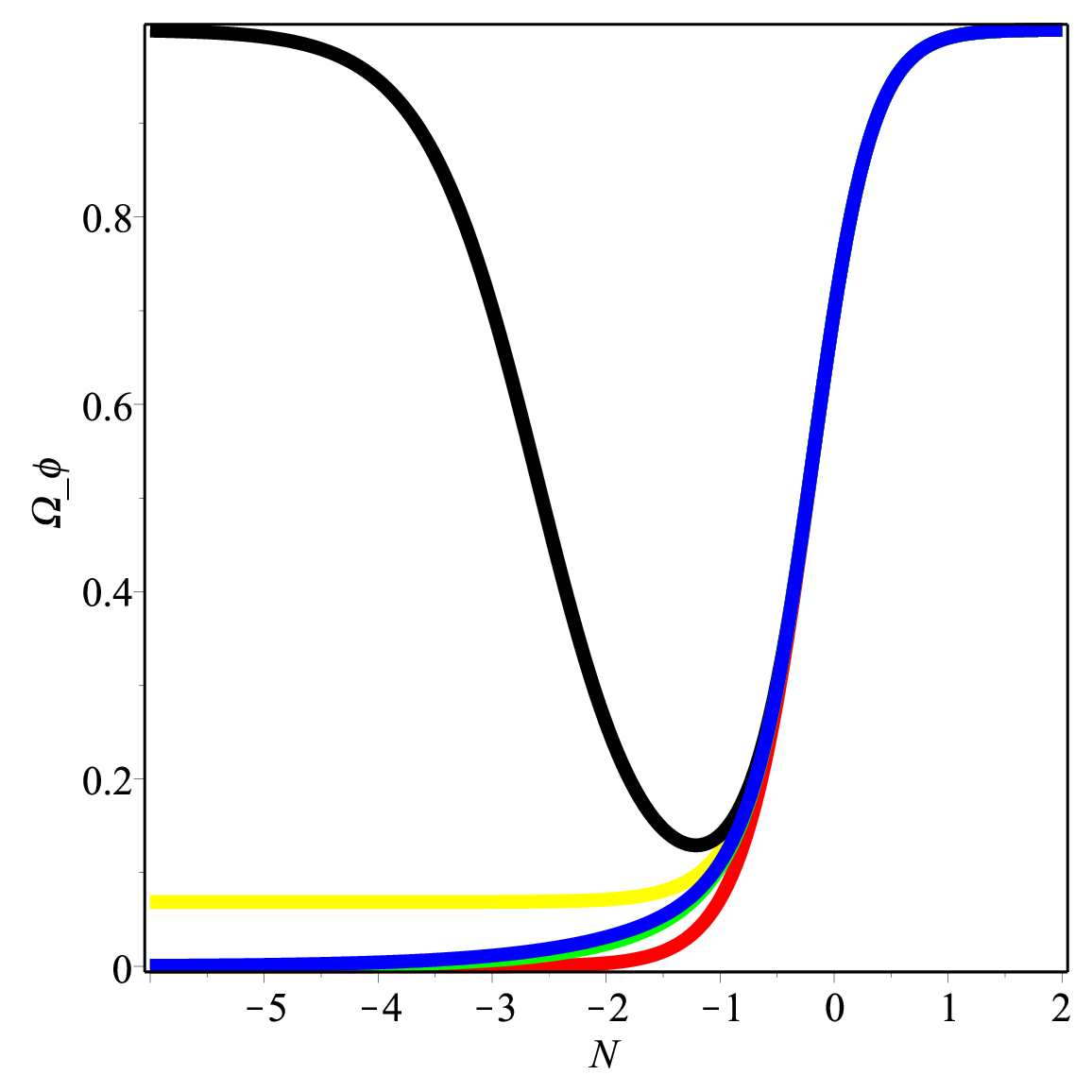}
 \caption{The evolution of $\Omega_{\phi}$ with respect to $N$
when $F(X)=A_1\sqrt{X}-A_2 X^{\beta}$.  The initial condition and
the values of the parameter $\eta$ are the same as Fig.\ref{fig9}
except $\gamma_b=4/3$. Fig.\ref{fig9} and Fig.\ref{fig10} are
plotted using Eq.(\ref{eqsadd18}).}\label{fig10}
\end{minipage}
\end{figure}

 For the cosmic evolution in very early time, $N$ is negative and $|N|$ is very large, we get $\gamma_{\phi}\approx \eta=2\beta/(2\beta-1)$
 from Eq.(\ref{eqsadd18}). The scalar field will mimic the evolution of matter with zero pressure $\rho\sim a^{-3}$ in the
 limit of $\beta\rightarrow\infty$. For the cosmic evolution in late time, $N$ is very large, $\gamma_{\phi}\approx 0$ and $\Omega_{\phi}\approx 1$,
 scalar field behaves as the cosmological constant. So it is the second case that a lagrangian without a potential term gives the cosmological constant
 solution. Moreover, we know from Eq.(\ref{eqsadd18}) that $\Omega_{\phi}$ might not be 0 in the very early time. It depends on the
 value of $\eta$ comparing with the value
 of $\gamma_b$($\gamma_b=1$ for matter and $\gamma_b=4/3$ for radiation). When $N \rightarrow -\infty$,  we can get
 $\Omega_{\phi}(N) \approx 1/(c_6 e^{3(\eta-\gamma_b)N}+1)$ from Eq.(\ref{eqsadd18}). Then $\Omega_{\phi}\rightarrow 1$ for $\eta>\gamma_b$,
 $\Omega_{\phi}\rightarrow 1/(c_6+1)$ for $\eta=\gamma_b$ and $\Omega_{\phi}\rightarrow 0$ for $\eta<\gamma_b$. In this case, it will be interesting to
 investigate the impact of the scalar field on the evolutional history of the early universe. We have plotted the evolution of $\Omega_{\phi}$ with
 respect to $N$ for different $\eta$ and different $\gamma_b$ in Fig.\ref{fig9} and Fig.\ref{fig10},
 and it is shown that the value of $\Omega_{\phi}$ in the very early time can actually be 1, 0 or an arbitrary positive constant which is less than 1.

\section{Cosmological Implications and Conclusion}
 \par The main purpose of the paper is not to analyze the dynamical behavior about different scalar fields in detail, so we will not investigate the
 detailed critical points and their stable properties for each dynamical system. What we want to focus on is about the dynamical system itself.

\subsection{Variables $(x, y)$ vs observable quantities $(\gamma_{\phi}, \Omega_{\phi})$}

 \par Dynamical variables ($x, y$) in the previous papers are about the scalar field and its first
 derivative. Though for some scalar field models(quintessence or phantom quintessence),
the combination of variables ($x, y$) has certain meaning, namely ,
$y^2 \pm x^2=\Omega_{\phi}$, but generally speaking, these dynamical
variables have no direct cosmological meaning, and the autonomous
system for $dx/dN$ and $dy/dN$ also varies from different models.
Authors studied dynamical properties of scalar field based on
 quantities $(\gamma_{\phi}, \Omega_{\phi})$ instead of variables $(x, y)$\cite{add2, add3, 9, 7, 8, 10, 12, 13,
 add1}. It is more convenient if we change the variables from $(x, y)$ to observable quantities $(\gamma_{\phi}, \Omega_{\phi})$.
 Firstly, $(\gamma_{\phi}, \Omega_{\phi})$ are directly related to
the observable quantities and also about the properties of dark
energy. Analyzing the system based on $(\gamma_{\phi},
\Omega_{\phi})$, we can figure out how the equation of state of dark
energy $w_{\phi}$ and the density
 parameter $\Omega_{\phi}$ evolve. Secondly, though the form of autonomous system for $dx/dN$ and $dy/dN$ are completely different for different models,
 but it is quite interesting that the function for $d\Omega_{\phi}/dN$ has the same expression in quintessence, tachyon, k-essence and general non-canonical scalar
  field model(i.e., Eq.(\ref{eqs14}),Eq.(\ref{eqs23}),Eq.(\ref{eqs35}),Eq.(\ref{eqs44})) as follows:

\begin{equation}\label{eqs47}\frac{d \Omega_{\phi}}{dN}=3(\gamma_b-\gamma_{\phi})\Omega_{\phi}(1-\Omega_{\phi})\end{equation}

 The only difference is the form of function $d\gamma_{\phi}/dN$. In fact, it is a well-known fact that Eq.(\ref{eqs47}) holds for all non-coupled
dark energy models as long as they satisfy the following equations:
\begin{equation}\label{eqs48} H^2=\frac{1}{3M^2_{pl}}(\rho_b+\rho_{de}), ~ \dot{\rho}_i+3H(p_i+\rho_i)=0\end{equation}

 where subscript $i$ denotes each energy component such as dark energy, matter or radiation. If we set $\gamma_{de}=w_{de}+1=p_{de}/\rho_{de}+1$ and
 $\Omega_{de}=\rho_{de}/3M^2_{pl}H^2$,  Eq.(\ref{eqs47}) can be
 derived from  Eq.(\ref{eqs48}). So the dark energy density
 parameter $\Omega_{\phi}$ obeys the same differential equation( namely
 Eq.(\ref{eqs47})) independent on the scalar field models considered
 whenever the dark energy is uncoupled in GR frame. For example, the authors got the same equation
 even for the purely kinetic coupled gravity model which modified the standard general relativity action through the addition of a coupling between
 functions of the metric and kinetic terms of a free scalar field\cite{24}(Eq.(28) in this paper
 and $\gamma_b$ is taken as 1).

 \par We can conclude from  Eq.(\ref{eqs47}) that there are only three possible destinies(three types of critical points) for $\Omega_{de}$ to be in these
 models, namely $\Omega_{de}=0$, $\Omega_{de}=1$ and the case $\gamma_{de}=\gamma_b$ where the value of $\Omega_{\phi}$ is determined by other equation in the
 dynamical system. The cases of $\Omega_{de}=0$ or $\Omega_{de}=1$ are completely opposite destinies, corresponding to the universe completely dominated only
 by the scalar field or by the barotropic fluid. Generally speaking, we can  obtain $0<\Omega_{de}<1$ for the case of $\gamma_{de}=\gamma_b$. However,
 for this scaling solution, the equation of state of dark energy
$w_{de}$ is the same as the equation of state of barotropic fluid
$w_b$, so there is no accelerating expansion. Since the observation
suggested that we are living in an accelerated expanding  universe
with $\Omega_{de}\sim0.7$, none of these three destinies correspond
to the present universe we observed. This could be
  considered as a clue for the possibility of the interaction between dark energy and other barotropic fluids(see \cite{25} for such
 model) if we want to solve or at least alleviate the cosmological coincidence problem
 without fine-tunings. This result is valid for not only all the non- coupled dark energy models, but also for many modified gravity models as long as the energy
 density and the pressure of dark energy or effective dark energy satisfy the continuity equation
 Eq.(\ref{eqs48}). However, we should emphasize  that our result is
 not new, there are many works on the study of interacting model of
 dark energy model\cite{add2, add3, add4, add5, add6}.

\subsection{Two-Dimensional vs Three-Dimensional dynamical autonomous system}

 \par Another important thing we want to emphasize is that, it is more reasonable and more scientific to investigate the dynamical behaviors of a dark
energy model under the three-dimensional autonomous system rather
than the two-dimensional system.  Firstly, the two-dimensional
dynamical autonomous system is just a specific case when the
potential takes a special form. if we want to completely study the
general dynamical properties of a dark energy model, we need to
study the system beyond a special potential. Then we can find more
critical points than the ones found in a two-dimensional system. We
therefore are able to analyze which critical points are possessed by
a class of dark energy models and which ones exist only due to the
concrete potentials. The method studying the three-dimensional
dynamical autonomous system beyond one special potential is
originated for the quintessence \cite{6, 26} and then developed to
other dark energy models \cite{cite61, cite62, cite63, cite65,
cite66, cite67, cite68, cite69, 25, cite611}. Here we extend this
method to the more general scalar field models in Section 3.
Secondly, more stable attractors can be found in terms of three
dimensional autonomous system. For example, a new critical point is
found only in three dimensional dynamical system of power-law
kinetic quintessence, which corresponds to the dark energy dominated
universe($\Omega_\phi=1$) where power-law kinetic quintessence
behaves as an cosmological constant with the sound speed $c_s^2$
being 0\cite{add7}. Thirdly, from the viewpoint of chaos theory, the
dynamical properties of three-dimensional autonomous system is more
fruitful than the two-dimensional system. According to the
Poincar$\acute{e}$ -Bendixson theorem, chaos does not exist in any
two-dimensional autonomous dynamical system\cite{27, 28} but could
be possible in three-dimensional autonomous dynamical system. For a
number of three-dimensional systems, such as the famous
three-dimensional Lorenz equations which is a model describing the
atmospheric convection \cite{29}, there exist chaos for certain
values of the parameters.

\subsection{Stable attractors vs chaotic behaviors}

\par The studies of chaotic dynamics in cosmological models has a
long story. Chaotic properties had reported in spatially closed
scalar field FRW cosmological models \cite{30, 31, 32, 33, 34, 35},
spatially flat FRW cosmological model with two or more scalar fields
\cite{36,37}, Bianchi IX universe \cite{38, 39}, Bianchi I
universe\cite{40} and the mixmaster universe\cite{41}. It would be
very interesting and also a big challenge for the theoretical study
of dark energy if the dynamical systems we consider here(i.e.,
Eqs.(\ref{eqs14}-\ref{eqs16}),Eqs.(\ref{eqs23}-\ref{eqs25}),Eqs.(\ref{eqs35}-\ref{eqs37})
and Eqs.(\ref{eqs44}-\ref{eqs46}) ) exist the chaotic
 properties. Then the evolution of $\Omega_{\phi}$ and $\gamma_{\phi}$
 will be very sensitive to the initial condition, and therefore predicting their evolution in future becomes totally
 impossible. However, it is proved that there is no chaotic behavior in spatially flat single scalar field FRW cosmological
 models\cite{42, 43}. Since for the spatially flat case with $k=0$, the dynamical system can be described by a three-dimensional autonomous system
 with a set of variables $(H, \phi, \dot\phi)$ under a Hamiltonian constraint, so the dynamical system is actually a two-dimensional autonomous system($a$ and
 $\dot a$ appear only in the combination $H=\dot a/ a$)\cite{44}. We know that for the two-dimensional autonomous systems, there are no enough degrees of freedom to exist chaos,
 so this proved no-chaotic dynamics in the spatially flat scalar field FRW cosmological model. However, we noted that this result is obtained in the
 absence of matter and radiation. This may be the case in the very early time when our universe is undergoing an inflation era and completely dominated
 by the scalar field. However, for the study of dark energy of late-time cosmic acceleration, the component of matter is comparable with the density of
 dark energy and should not be ignored when we investigate the dynamical behavior of scalar field. In the presence of matter, scale factor
 $a$ will reappear in the dynamical system beside the variables $(H, \phi, \dot\phi)$, and then the system can not be reduced to two-dimensional
 autonomous dynamical system any more. So here we argue that, beside the ordinary attractors(such as dark energy dominated solution,
 de-Sitter like solution and scaling solution), it is still possible for
the chaotic behavior in spatially flat single scalar field FRW
cosmological
 models in the presence of matter. It is very like the case of spatially non-flat($k\neq 0$) single scalar field FRW cosmological models where the
 dynamical system can not be reduced to two-dimensional autonomous dynamical
 system too. What we argued here is also supported by the equations in section 3( i.e., Eqs.(\ref{eqs14}-\ref{eqs16}), Eqs.(\ref{eqs23}-\ref{eqs25}),
 Eqs.(\ref{eqs35}-\ref{eqs37}) and Eqs.(\ref{eqs44}-\ref{eqs46})), which described three-dimensional autonomous dynamical systems. However,
 we are not sure whether there truly exist the chaotic behavior in spatially flat scalar field FRW cosmological models
 now, to find the chaotic behavior is beyond the scope of this paper, it should be investigated in future.

\section{Acknowledgement}
\par We owes the great improvement of this paper to the anonymous referees. This work is partly supported by Chinese National Nature
Science Foundation under Grant No.11333001, No.11433003, 973 Program
No.2014CB845704 and Shanghai Science Foundations 13JC1404400. This
work is also supported by Shanghai Normal University.


\begin{thebibliography}{0}    

\bibitem{1} B. Ratra and P. J. E. Peebles, Phys. Rev. D\textbf{37}, 3406 (1988)
\bibitem{2} E. J. Copeland, M. Sami and S. Tsujikawa, Int. J. Mod. Phys. D\textbf{15}, 1753 (2006)
\bibitem{3} M. Li, X. D. Li, S. Wang and Y. Wang, Commun. Theor. Phys\textbf{56}, 525-604 (2011)
\bibitem{add2} A. A. Coley, \cal $Dynamical ~Systems ~and ~Cosmology, ~Kluwer ~Academica~Publishers (2003)$
\bibitem{Saridakis1} C. Xu, E. N. Saridakis and G. Leon, JCAP \textbf{0904}, 001 (2009)
\bibitem{Saridakis2} G. Leon and E. N. Saridakis, JCAP\textbf{1303}, 025 (2013)
\bibitem{Saridakis3} G. Leon, J. Saavedra and E. N. Saridakis, Class.Quant.Grav.\textbf{30}, 135001  (2013)
\bibitem{Saridakis4} C. R. Fadragas, G. Leon and E. N. Saridakis, Class. Quantum Grav.\textbf{31}, 075018 (2014)
\bibitem{4} J. M. Aguirregabiria, L. P. Chimento and R. Lazkoz, Phys. Lett. B\textbf{631}, 93-99 (2005)
\bibitem{5} A. R. Liddle, Paul Parsons and J. D. Barrow, Phys. Rev. D\textbf{50}, 7222-7232 (1994)
\bibitem{6} W. Fang, Y. Li, K. Zhang and H. Q. Lu, Class. Quantum Grav.\textbf{26}, 155005 (2009)
\bibitem{cite61} I. Quiros, T. Gonzalez, D. Gonzalez, Y. Napoles, R. Garcia-Salcedo and C. Moreno,  Class. Quant. Grav.\textbf{27}, 215021(2010)
\bibitem{cite62} Y. Leyva, D. Gonzalez, T.Gonzalez, T. Matos and I. Quiros, Phys. Rev. D\textbf{80}, 044026 (2009)
\bibitem{cite63} T. Matos, J. R. Luevano, I. Quiros, L. A. Urena-Lopez and J. A. Vazquez, Phys. Rev. D\textbf{80}, 123521 (2009)
\bibitem{cite65} K. Xiao and J. Y. Zhu, Phys. Rev. D\textbf{83}, 083501 (2011)
\bibitem{cite66} D. Escobar, C. R. Fadragas, G. Leon and Y. Leyva, Class. Quantum Grav.\textbf{29}, 175005 (2012)
\bibitem{cite67} D. Escobar, C. R. Fadragas, G. Leon and Y. Leyva, Class. Quantum Grav.\textbf{29}, 175006 (2012)
\bibitem{cite68} G. Leon, Y. Leyva and J. Socorro, arXiv:1208.0061
\bibitem{cite69} S. del Campo, C. R. Fadragas, R. Herrera, C. Leiva, G. Leon and J. Saavedra, Phys. Rev. D\textbf{88}, 023532 (2013)
\bibitem{cite611} G. Otalora, Phys. Rev. D\textbf{88}, 063505 (2013)
\bibitem{add2} L. P. Chimento, A. S. Jakubi and D. Pavon, Phys. Rev. D\textbf{67}, 087302 (2003)
\bibitem{add3} L. P. Chimento, A. S. Jakubi, D. Pavon and W. Zimdahl, Phys. Rev. D\textbf{67}, 083513 (2003)
\bibitem{9} J. G. Hao and X. Z. Li, Phys. Rev. D\textbf{70}, 043529 (2004)
\bibitem{7} R. J. Scherrer and A. A. Sen, Phys. Rev. D\textbf{77}, 083515 (2008)
\bibitem{8} G. Gupta, S. Majumdar and A. A. Sen, MNRAS\textbf{420}, 1309 (2012)
\bibitem{10} S. Dutta and R. J. Scherrer, Phys. Lett.B\textbf{704}, 265-269 (2011)
\bibitem{12} W. Fang and H. Q. Lu, Eur. Phys. J. C\textbf{68}, 567-572 (2010)
\bibitem{13} N. C. Devi, T. R. Choudhury and A. A. Sen, MNRAS\textbf{432}, 1513-1524 (2013)
\bibitem{13add} X. M. Chen and Y. G. Gong, arXiv:1309.2044
\bibitem{add1} Y. G. Gong, arXiv:1401.1959
\bibitem{add4} L. Amendola, Phys. Rev. D\textbf{62}, 043511 (2000)
\bibitem{add5} R. Curbelo, T. Gonzalez, Genly Leon and I. Quiros, Class. Quant. Grav. \textbf{23}, 1587 (2006)
\bibitem{add6} T. Gonzalez, G. Leon and I. Quiros, Class. Quant. Grav. \textbf{23}, 3165 (2006)
\bibitem{14} R. J. Yang and X. T. Gao, Chin. Phys. Lett.\textbf{26}, 089501 (2009)
\bibitem{Tamanini2014} N. Tamanini, arXiv:1401.6339
\bibitem{16} T. Chiba, T. Okabe and M. Yamaguchi, Phys. Rev. D\textbf{62}, 023511 (2000)
\bibitem{17} R. J. Yang and X. T. Gao, Class. Quant. Grav.\textbf{28}, 065012 (2011)
\bibitem{18} J. Garriga and V.F. Mukhanov, Phys. Lett. B\textbf{458}, 219-225 (1999)
\bibitem{19} A. Vikman, Phys. Rev. D\textbf{71}, 023515 (2005)
\bibitem{20} J. De-Santiago, J. L. Cervantes-Cota and D. Wands, Phys. Rev. D\textbf{87}, 023502(2013)
\bibitem{21} R. J. Scherrer, Phys. Rev. Lett.\textbf{93}, 011301 (2004)
\bibitem{22} J. L. Cervantes-Cota, A. Aviles and J. De-Santiago, AIP Conf. Proc.\textbf{1548}, 299-313 (2013)
\bibitem{23} L. P. Chimento, Phys. Rev. D\textbf{69},  123517 (2004)
\bibitem{24} G. Gubitosia and E. V.Linder, Phy. Lett. B\textbf{703}, 113-118 (2011)
\bibitem{25} G. Otalora, JCAP\textbf{07}, 044 (2013)
\bibitem{26} S. Y. Zhou, Phys. Lett. B\textbf{660}, 7-12(2008)
\bibitem{27} S. Wiggins,  \cal $Introduction ~to ~Applied ~Nonlinear ~Dynamical ~Systems ~and ~Chaos$ (New York: Springer, 1990)
\bibitem{28} F. Zhang and J. Heidelz, Nonlinearity\textbf{10},  1289-1303(1997)
\bibitem{29} E. N. Lorenz, Journal of Atmospheric Sciences\textbf{20}, 130(1963)
\bibitem{30} S. Blanco, G. Domenech, C. El Hasi, and O. A. Rosso, Gen.Rel.Grav.\textbf{26}, 1131-1143 (1994)
\bibitem{31} E. Calzetta and C. El Hasi, Class. Quant. Grav.\textbf{10}, 1825 (1993)
\bibitem{32} E. Calzetta and C. El Hasi, Phys. Rev. D\textbf{51}, 2713(1995)
\bibitem{33} A. V. Toporensky, Int.J.Mod.Phys. D\textbf{8}, 739-750 (1999)
\bibitem{34} S. E. Jor$\acute{a}$s and T. J. Stuchi, Phys.Rev. D\textbf{68}, 123525 (2003)
\bibitem{35} G. Lukes-Gerakopoulos, S. Basilakos and G. Contopoulos, Phys. Rev. D\textbf{77}, 043521 (2008)
\bibitem{36} N. J.Cornish and J. J. Levin, Phys. Rev. D\textbf{53}, 3022-3032 (1996)
\bibitem{37} R. Easther and Kei-ichi Maeda, Class. Quant. Grav.\textbf{16}, 1637-1652 (1999)
\bibitem{38} S. Fay and T. Lehner, Gen.Rel.Grav.\textbf{37}, 1097-1117 (2005);
\bibitem{39} E. J. Kim and S. Kawai,Phys. Rev. D\textbf{87}, 083517 (2013)
\bibitem{40} J. H. Chen and Y. J. Wang, Chinese Phys.\textbf{14}, 1282 (2005)
\bibitem{41} N. J. Cornish and J. J. Levin, Phys. Rev. Lett.\textbf{78}, 998-1001 (1997)
\bibitem{42} E. Gunzig, L. Brenig, A. Figueiredo and T.M. Rocha Filho, Mod. Phys. Lett. A\textbf{15}, 1363-1368 (2000)
\bibitem{43} V. Faraoni, M. N. Jensen and S. A. Theuerkauf, Class.Quant.Grav.\textbf{23}, 4215-4230 (2006)
\bibitem{44} O. Hrycyna, \cal $Regular ~and ~Chaotic ~Dynamics ~in ~Scalar ~Field ~Cosmology$, doctoral dissertation (2011)
\bibitem{add7} W. Fang, H. Tu, Y. Li, J. S. Huang and C. G. Shu, Phy. Rev. D\textbf{89}, 123514 (2014)
\end{thebibliography}
\end{document}